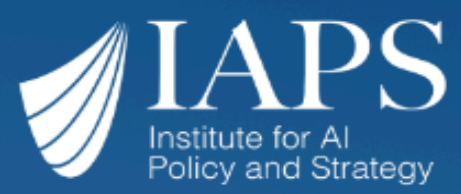

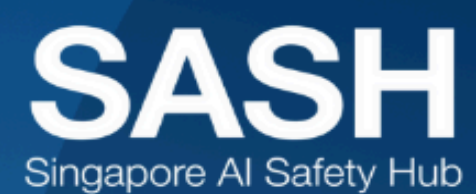

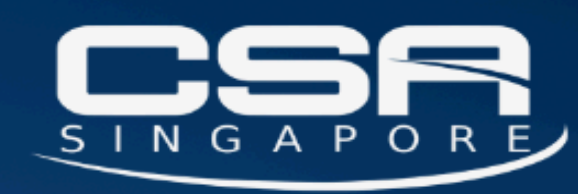

MAY 2026

# Detecting Offensive Cyber Agents

A Detection-in-Depth Approach

**AUTHORS**

Matthew Mittelsteadt, Jam Kraprayoon, Robin Staes-Polet, Oskar Galeev, Jan Wehner, Christopher Covino, Shaun Ee

## AUTHORS

Matthew Mittelsteadt*[1]

Jam Kraprayoon*[1]

Robin Staes-Polet†

Oskar Galeev‡

Jan Wehner*

Christopher Covino*

Shaun Ee*

## ORGANIZATIONS

*Institute for AI Policy and Strategy

†Existential Risk Alliance

‡Singapore AI Safety Hub

---

[1] Research lead(s)

# Table of Contents

# Executive Summary

Artificial Intelligence (AI) agents can now orchestrate cyberattacks. In September 2025, Anthropic detected a Chinese state-sponsored threat actor, designated GTG-1002, that used Claude Code to simultaneously attack around thirty organizations. AI agents automated approximately 80–90% of the operation. Automated attack capabilities are also visible in evaluation settings: In April 2026, the UK AI Security Institute found that both Claude Mythos Preview and GPT-5.5 could complete multi-step simulated network intrusions, end-to-end, without human intervention.

Automated attack capabilities would dramatically alter the nature of cyber threats. As offensive AI continues to advance, agentic attacks could compress operational timelines, scale the number of simultaneous operations far beyond human limits, lower the cost and expertise required for sophisticated intrusions, and adapt independently in ways that conventional defenses are not built to handle.

## Detection Must Be a Strategic Priority

Preparing for these threats requires progress across four interlocking defensive layers: *Delay* (slowing high-risk cyber capability proliferation), *Defend* (hardening targets and reducing attack surfaces), *Detect* (identifying offensive cyber agent activity), and *Disrupt* (imposing costs and neutralizing offensive cyber agents). Action across each layer is essential to creating the resilience needed to stand up to emerging threats. **This report argues, however, that the *Detect* layer should be elevated as a strategic priority**.

Robust detection capabilities can unlock significant strategic benefits. **Detection is a key enabler for other countermeasures like defense and disruption**. Triaging intrusions, adapting defenses, and disrupting malicious infrastructure all depend on first identifying and understanding the offensive cyber agent threat. **Detection also underpins situational awareness.** Policymakers currently lack an empirical basis for assessing this threat landscape—only through strong detection can this challenge be quantified, tracked, and assessed. Unfortunately, **our understanding of offensive cyber agent detection remains deeply underdeveloped**. These capabilities have only just emerged, and knowledge of their signatures, behavior, and strategic patterns is almost entirely lacking. Immediate research is needed to develop the policy and technical mechanisms required for an agent-ready detection approach.

## The Detection Challenge

Detecting offensive cyber agents presents three major challenges:

1. **Highly adaptive agents may evade existing detection methods**: We expect offensive cyber agents to leverage high operational and behavioral variance, making detectable signatures harder to establish and shorter-lived than conventional attacks. Agents may evade traditional detection methods through novel operations, dynamic code generation, and deliberate behavioral and tactical variation. Unlike human attackers, agents may also operate below anomaly-detection thresholds and blend in with legitimate agent tools deployed in network environments. For these reasons, a significant **detection gap** between conventional and agentic attacks appears likely.
2. **Agent-orchestrated attacks may distribute actions across multiple models, vendors, and platforms**: Agent-enabled attacks are already disaggregating attack steps and requests across models and systems. If this trend continues, any single point of observation will be unable to aggregate the data sufficient for robust detection.
3. **Offensive cyber agents may horizontally scale highly sophisticated attacks yielding *everything, everywhere, all-at-once*-style campaigns**: High sophistication attacks once constrained by human limits may soon be bottlenecked primarily by processing power. This shift will enable campaigns that strike multiple targets and sectors simultaneously. This risks straining detection resources and threat triage capabilities. Existing cyber infrastructure built around sectoral and national "silos" may be unfit for such systemic, cross-sectoral attack breadth.

## Detection-in-Depth: A Strategic Approach

This report proposes a strategic framework called detection-in-depth to guide forward action by defenders and policymakers seeking to address these challenges. **Detection-in-depth rests on a key structural insight: research shows even modest improvements in detection capability could sharply reduce the likelihood of attacker success—yet no single detection mechanism, process, or actor will be sufficient to detect offensive cyber agent threats.** Agentic detection, therefore, requires defenders to both prioritize detection and implement layers of technical mechanisms to enable visibility not only within individual networks, but across the ecosystem. Detection-in-depth is organized around several core functions: **Discovery**, or generating novel signals that make agentic activity discoverable, **Analysis**, or interpreting those signals to confirm malicious activity and produce actionable intelligence, and **Coordination**, or uniting organizations to pool dispersed signals to enable ecosystem-wide threat awareness.

## Detection Mechanisms

To operationalize detection-in-depth, this report proposes three classes of detection mechanisms:

**Identity mechanisms** operate at the access level, before or at the point where an agent enters a system. By tracking, establishing, verifying, and logging digital actors, these mechanisms help answer who an digital actor is, what its intentions are, and whether its activity can be tracked. As a starting point, we recommend **Agent Identifiers for Critical Infrastructure**: persistent, cryptographically verifiable credentials attached to agent traffic that generate detection-relevant telemetry when interacting with critical infrastructure operators. These identifiers aid detection by generating telemetry about agent activity where almost none currently exists.

**Environmental detection mechanisms** operate at the activity level, once a digital actor is already present inside a network. They observe what the digital actor does rather than who it claims to be. We recommend two specific mechanisms. **Agent Honeypots** are decoy systems designed specifically to attract autonomous attackers. Honeypots can extract threat intelligence about how malicious agents operate in real-world conditions. **AI-Automated Alert Analysis and Triage** systems use AI to filter, prioritize, and interpret the growing volume of detection signals expected from autonomous cyber operations.

**Information sharing and analysis mechanisms** operate at the ecosystem level, aggregating, correlating, and sharing signals across organizations to surface patterns that no single actor could detect alone. We recommend two mechanisms. Agent service providers can adopt an **Agentic Security Alert Standard** to improve the speed, consistency, and actionability of agent threat reporting across organizations. Industry and governments can establish an **Agentic Cybersecurity Exchange (ACE),** a global information-sharing and disruption coordination institution modeled on the Global Signal Exchange. The ACE is specifically tailored to aggregate and analyze signals from model providers, cloud platforms, and other agent infrastructure operators about offensive cyber agent threats to aid detection and coordinate threat disruption.

## Recommendations

This report proposes 12 recommendations to support detecting offensive cyber agents:

| Mechanism | Recommendation | Responsible Actor(s) |
|---|---|---|
| **Agent Identifiers for Critical Infrastructure** | Negotiate convergence on interoperable agent identity standards before fragmented proprietary standards harden | • Governments<br>• Standards bodies<br>• Agent service providers<br>• Critical infrastructure owners and operators |
| | Require and/or incentivize Agent ID adoption among critical infrastructure owners and operators | • Governments |
| | Prepare institutional infrastructure for cross-organizational agent identity correlation | • Governments |
| | Invest in privacy-preserving technology and more robust runtime attestation to support agent identity infrastructure | ● Governments<br>● Research funders |
| **AI-Automated Alert Analysis and Triage** | AI companies with advanced cyber-capable models should work with defenders and security researchers to develop and test AI-automated alert analysis and triage tools | ● AI companies<br>● Defenders<br>● Security researchers |
| **Agent Honeypots** | Fund public research and development into agent honeypot design and detection techniques | ● Governments<br>● Philanthropists |
| | Existing honeypot operators should begin collecting data on agent activity | ● Existing honeypot operators |
| | Create market incentives for private-sector development of agent honeypot products | ● Governments<br>● Philanthropists |
| **Agent Security Alert Standard** | Lead agent security alert standards development | ● AI industry consortia or trusted independent and national standards bodies |
| **Agentic Cybersecurity Exchange (ACE)** | Develop ACE implementation plans | ● Governments<br>● Agent service providers (AI companies, cloud providers, other agent ecosystem service providers) |
| | Assess the feasibility of a CRISP-like sensing architecture tailored to agentic threat detection | ● Governments<br>● Agent service providers |
| | Invest in ecosystem-wide information sharing and analysis research | ● Philanthropists and research funders<br>● Industry<br>● Governments |

# 1 | Introduction

In June 2025, the Computer Emergency Response Team of Ukraine (CERT-UA) documented Russian state-sponsored actors deploying a new kind of weapon against Ukrainian targets.[2] LAMEHUG did not rely on pre-written malicious code. Instead, it queried a large language model (LLM) in real time, generating attack commands on the fly. This was not an isolated incident. Across several cases,[3] attackers have started to deploy novel AI-enabled malware in active operations.

These cases were only a prelude. In September 2025, Anthropic detected what it described as something more significant still: a threat actor, which they assessed with high confidence to be a Chinese state-sponsored group, used Anthropic's Claude Code agent to attempt infiltration into roughly 30 organizations across tech, finance, government, and critical infrastructure. What distinguished this operation was not its targets but its architecture: AI agents performed 80 to 90% of the campaign autonomously, with minimal human involvement.[4] Meanwhile in February 2026, researchers discovered a compromise of Mexican government and water utilities, in this case enabled by the pairing of OpenAI GPT class models with Claude.[5]

These real world successes are now visible in evaluation settings: in April 2026, the United Kingdom AI Security Institute (UKAISI) found that both Claude Mythos Preview and GPT-5.5 can complete simulated 32-step network intrusions end-to-end.[6] While the UKAISI report stresses this only demonstrates success against small, weakly defended networks, this success should be treated as a harbinger for future developments to come.

These demonstrated end-to-end attack capabilities sit against a backdrop of rapidly accelerating benchmark performance. In 2023, the UKAISI found that AI systems could barely complete apprentice-level cyber tasks in 2023.[7] Today, similar models not only complete close to 100% of apprentice-level tasks but up to 73% of expert-level tasks—those that previously required the equivalent of a decade of expert human experience.[8] Irregular, a security-focused evaluation organization, found that cybersecurity techniques once associated primarily with top-tier state actors and expert researchers, such as exploiting flaws in cryptographic signature schemes, are starting to be accomplished by frontier AI systems.[9]

---

[2] CERT-UA, "UAC-0001 cyberattacks on the security and defense sector using the LAMEHUG software tool, which uses LLM (large language model) (CERT-UA#16039)."

[3] For instance, in the s1ngularity attack in September 2025 attackers hijacked targets' agents to manage data exfiltration. See Winterford, "The s1ngularity attack: When attackers prompt your AI agents to do their bidding."

[4] Hallucinations are a function of an as-yet immature technology. Recent evidence suggests, however, that this challenge is rapidly fading. In 2025, bug bounty programs were plagued by a massive influx of AI-generated, hallucination-laden bug reports. In 2026, this problem appears to be gone. According to Curl, despite the continued prevalence of AI-generated bug reports, the number of reports containing confirmed vulnerabilities is now higher than in the pre-AI era. See Stenberg, "High-Quality Chaos."

[5] Dragos, "AI in the Breach: How an Adversary Leveraged AI to Target a Water Utility's OT"

[6] AI Security Institute, "Our Evaluation of OpenAI's GPT-5.5 Cyber Capabilities."

[7] D'Cruz et al., "AISI Frontier AI Trends Report (2025)."

[8] AI Security Institute, "Our Evaluation of OpenAI's GPT-5.5 Cyber Capabilities."

These early signals around the use of autonomous agents to conduct cyber campaigns suggest that the cost and complexity of sophisticated cyber operations may be entering a period of rapid, sustained decline, with effects on the balance between attacker and defender that are only starting to become clear. This trajectory points toward a world where expert-level cyber operations are available "in the cloud" to any attacker, rentable by the hour. Even more concerning is a potential future where sophisticated attacks can be launched from an open-source model running on private infrastructure, beyond the reach of any model provider or cloud host.

The emergence and proliferation of autonomous offensive cyber operations introduce four concerning threat actor dynamics.

**Speed:** Autonomous agents could dramatically accelerate nearly every phase of a cyber campaign.[10] Preparing a sophisticated operation against a hardened target currently takes months: researching the target, building infrastructure, and developing tailored exploits. Maintaining a foothold once inside can take months more. Agents, acting with machine speed and parallelized at scale, could compress times involved with preparation, often the lengthiest part of the cyber kill-chain. The same logic applies to every stage that follows, and the windows defenders rely on for detection and response shrink accordingly. Any traditional defenses that rely on humans for analysis (e.g., patching, auditing) may soon be rendered ineffective.

**Scale:** Human threat actors can only manage so many simultaneous intrusions, forcing even well-resourced groups to be selective about their target set. Cyber agents could pursue thousands of targets simultaneously, tailoring attacks to each and managing intrusions in parallel.[11] Such operational scale could overwhelm defensive resources—flooding threat information sharing systems, straining limited law enforcement capacity, and further taxing defenders already facing systemic fatigue.

**Cost:** The need for specialized talent for more sophisticated cyber operations has historically been a major constraint on how many actors could conduct them. If AI inference becomes cheaper and capable systems more widely accessible, operations that once required nation-state resources become available to criminal groups, and operations that once required criminal groups become available to individuals.[12] With hacking democratization, defenders will face not only more attacks but potentially attacks of higher sophistication.

---

[9] Irregular, "Evaluating GPT-5.2 Thinking: Cryptographic Challenge Case Study."
[10] Early adopters are already proving the power of AI speed. In 2025, Palo Alto found the exfiltration speed of the fastest quartile of attacks was quadruple the speed just one year prior, see Unit 42. See Palo Alto Networks, "Global Incident Response Report 2026."
[11] Lohn, "Defending Against Intelligent Attackers at Large Scales."
[12] Kraprayoon et al., "Highly Autonomous Cyber-Capable Agents."

**Operational autonomy:** Perhaps the most significant change is the shift from tools that execute instructions to systems that can reason and make decisions independently. Conventional automated attacks are correlated: botnets and worms run the same playbook across every infected node, meaning a single patch or shared signature can neutralize all attacks at once. Autonomous agents that adapt at each step break the assumption, eroding the structural advantages of layered defense[13] while enabling qualitatively different types of operations that may be hard for defenders to defeat. For example:

- AI-powered malware could overcome traditional constraints by operating inside victim networks with limited or even non-existent command-and-control,[14] dynamically changing tactics and rewriting its own code to evade detection, and processing data locally for stealthier exfiltration.[15]

- Rather than exfiltrating data in bulk, agents could analyze intelligence locally and transmit only what matters, increasing the value of these operations.[16]

- Sophisticated attacks have historically depended on zero-day exploits in widely-used software, such as the NSA-developed EternalBlue exploit for Microsoft Windows, which are capabilities so costly to develop that they are typically the preserve of nation-state actors. Autonomous agents could change this by conducting vulnerability research and exploitation without direct human guidance, probing a target's specific technology stack in real time, or chaining together smaller, known vulnerabilities to achieve equivalent effects.

Together, these dynamics reshape the operational environment in which cyber defense operates and create new detection challenges.

## Scope and Goals of This Report

Together, these dynamics describe a potential period of acute risk. Offensive adoption of agentic tools could create a window of heightened cyber risk whose duration will depend on how quickly the defense community responds. The core problem this report addresses is, therefore: **Given the**

---

[13] Lohn, "Defending Against Intelligent Attackers at Large Scales."

[14] Harley "LOLMIL: Living Off the Land Models and Inference Libraries."

[15] As Crowdstrike 2025 threat hunting report assesses, traditional defenses and tools designed against single-domain attacks and non-autonomous malware can soon become obsolete, see Crowdstrike, "CrowdStrike 2025 Threat Hunting Report."

[16] The SparkCat malware campaign used on-device optical character recognition to scan victims' photo galleries and exfiltrate only images containing cryptocurrency wallet recovery phrases, see Kaspersky Team, "SparkCat trojan stealer infiltrates App Store and Google Play, steals data from photos." LLM-integrated malware could extend this further, filtering and summarizing a wide variety of data types by intelligence value before exfiltration, see Check Point Research, "AI in the Middle: Turning Web-Based AI Services into C2 Proxies & The Future of AI Driven Attacks."

**pace of offensive AI progress and the novel dynamics it introduces, how should defenses adapt to autonomous cyber operations?**

To mitigate risks from autonomous offensive cyber operations, defenders can pursue four interlocking strategic objectives[17]:

- **Delay:** Slowing the proliferation of high-risk AI tools that could enable malicious agentic capabilities
- **Defend:** Hardening targets and reducing the potential attack surface for malicious agents
- **Detect:** Identifying malicious agentic activity and enabling threat-environment visibility
- **Disrupt:** Imposing costs, actively degrading, and ultimately neutralizing malicious cyber-enabled agents

Action across each layer is essential for creating the resilience needed to stand up to emerging threats. Recent efforts such as Project Glasswing, a project aiming to *Delay* the proliferation of high-risk cyber capabilities through managed model access and *Defend* by targeting AI-assisted vulnerability discovery at key digital infrastructure, are commendable and must be built on. Yet, given the increasingly narrow window available to prepare for this threat, **we believe policymakers and firms must elevate detection as a strategic priority**. Our reasoning is threefold:

**Detecting autonomous threats can help enable other ways to counter offensive cyber agents, i.e., defense and disruption**. While some defensive measures, such as secure coding practices, operate independently, other measures, such as triaging intrusions and adapting defenses to emerging attacks, depend on first identifying and understanding the threat. Detection also supports disruption; for example, malicious agent infrastructure, such as inference compute or command-and-control channels, cannot be targeted without first being identified.

**Our understanding of agent detection is underdeveloped.** As cyber-capable agents have only just emerged, the study of their signatures, behavior, and impact is almost completely lacking. Further, research has yet to study how detection strategies must update in response to agentic dynamics. Detection strategy and mechanisms must ultimately seek to answer a range of questions that current approaches and infrastructure are poorly equipped to address—beginning with six that are foundational:

1. What activity is agentic, and what activity is human-directed?
2. Is this agent operating within its authorized boundaries?
3. Is this agent's behavior malicious?

---

[17] This framework was first described in Kraprayoon et al., "Highly Autonomous Cyber-Capable Agents."

4. What is the nature of this attack—what tactics, techniques, and procedures is it employing? What is the operational setup—what models and system components are employed?
5. Is this an isolated incident or part of a coordinated campaign?
6. Who is responsible, and under whose authority is this agent acting?

A robust detection approach capable of answering these questions cannot be developed overnight, so it is essential that we begin investigating both policies and technical mechanisms to underpin it as early as possible.

**Detection underpins situational awareness.** Policymakers currently have almost no empirical basis for assessing whether and how autonomous cyber operations are shaping the threat landscape. More comprehensive detection infrastructure would allow governments and defenders to gauge the scale of autonomous cyber activity, track how capabilities are proliferating across actors, assess what damage is occurring, and identify which actors are deploying these systems. Without this picture, policymakers cannot determine how much to prioritize the threat or where to allocate resources.

This report's goal is to **frame** both the challenges of agentic detection and a strategic approach to address them, while presenting **actionable mechanisms** to support policymakers, industry, and defenders in putting this strategy into practice.

In the sections that follow, we first outline the core challenges agentic detection will face. Next, we introduce detection-in-depth*,* our recommended strategic approach to agentic detection. Finally, we outline several promising mechanisms to put this approach into practice.

## Challenges of Detecting Autonomous Cyber Operations

The underdevelopment of agent detection should concern policymakers not only because the threat itself could be significant, but because detection of this kind will be uniquely challenging. We believe three core challenges confront any future detection effort:

1. **Agents could adapt behavior and operations enough to evade existing detection methods.**

While early research suggests agentic operations may yield some detectable signatures,[18][19] **we expect those signatures will be both harder to establish and shorter-lived than those of conventional attacks.** Autonomous agents combine the adaptiveness of human operators with the scale of automated tools, and may avoid developing the consistent tradecraft patterns that defenders learn to profile. Detection methods accumulate signal by observing repeated patterns. The higher the operational variance of an attacker, the lower the probability that any individual action crosses detection thresholds or recurs with sufficient frequency to be profiled. Agents may sustain operational variance high enough that the signal-to-noise ratio remains below defenders' detection thresholds. For this reason, a significant **detection gap** between conventional and agentic attacks appears likely.

[18] Zhang et al., "Exposing LLM User Privacy via Traffic Fingerprint Analysis: A Study of Privacy Risks in LLM Agent Interactions."
[19] Ghaleb, "Fingerprinting AI Coding Agents on GitHub."

**This will challenge the efficacy of the current cyber detection stack.** The most exposed contemporary detection methods are traditional signature-based techniques, which identify attacks by matching observed activity against known indicators of compromise (IOCs),[20][21] network payload signatures, YARA rules, and heuristic rules tied to known attacker tradecraft. These methods require threats to have been previously observed and to present recognizable signatures. Cyber offensive agents could undermine both conditions: through novel operations that have no prior equivalent, and by dynamically varying their code, tactics, and behavior.

Signature evasion is not new,[22] but it has historically been reserved for specialist cyber capabilities. Agentic tools could change this. **Through just-in-time code generation, advanced signature evasion will increasingly become a default malware property**. Evidence lends credence to this prediction: in 2025, Google's Threat Intelligence Group (GTIG) documented the first malware families that call out to LLMs during execution. PROMPTFLUX, one variant, interacts with the Gemini API and leverages a prompt instructing the LLM to rewrite the malware's entire source code on an hourly basis.[23]

In addition to signature-based methods, autonomous agents may also undermine behavioral analytics and Endpoint Detection and Response (EDR). For example, fuzzy techniques flag attacks through deviations from established network behavioral baselines. These methods have proven effective against living-off-the-land attacks, where adversaries use legitimate system tools rather than custom malware,[24] as human operators develop consistent tradecraft patterns[25] that can be profiled and detected over time. Agents could undermine these methods by systematically operating below anomaly-detection thresholds[26] and by blending into network environments, where they can use the same legitimate tools and APIs as the growing volume of authorized AI agents that organizations deploy for IT administration. Further, signatures may frequently shift over time as the underlying models, scaffolding, and tooling evolve.

---

[20] Example indicators include file hashes, IP addresses, domain names, and byte patterns.
[21] To illustrate, conventional cyber campaigns are often compromised once defenders identify code shared across multiple seemingly independent attacks. In the case of Stuxnet, Duqu, and Flame, reuse of code exposed connections between the campaigns. Source: Schwartz, "Flame Malware's Ties To Stuxnet, Duqu: Details Emerge."
[22] Polymorphic and metamorphic malware, which alter their own code with each new infection, have long exploited signature-based detection's dependency on consistency.
[23] Google Threat Intelligence Group (GTIG), "GTIG AI Threat Tracker: Advances in Threat Actor Usage of AI Tools."
[24] CISA et al., "PRC State-Sponsored Actors Compromise and Maintain Persistent Access to U.S. Critical Infrastructure."
[25] Living-off-the-land refers to a cyberattack in which attackers use legitimate software and functions available in the system to perform malicious actions on it, see Jarvis, "Leveraging Behavioral Analysis to Catch Living-Off-the-Land Attacks."

*Defensive Implication: Offensive Cyber Agents could adapt behavior enough to erode the signature and behavioral foundations of today's detection stack.*

2. **Agent-orchestrated attacks may distribute actions across multiple models, vendors, and platforms.**

The GTG-1002 campaign's ultimate detection and disruption demonstrated a weakness of orchestrating agentic attacks under one proprietary roof.[27] As agentic attack methods mature, future operations are likely to disaggregate across multiple platforms, models, cloud vendors, and toolsets, making any single point of observation insufficient for detection.

This trend is increasingly visible. First, individual steps in a multi-step attack can often be treated as discrete tasks and routed through separate model providers.[28] Such decomposition is not only feasible, but some evidence suggests it can indeed help skirt misuse detection: one study found that decomposing malicious code generation across Llama 70B and Claude 3 Opus raised the success rate to 43% compared to just 3% when models are used in isolation.[29] Already, such attack decomposition has been seen in multiple in-the-wild attacks,[30] including a campaign that combined Claude Code and GPT-4.1 to attack the Mexican government.[31]

Second, while these attack chains often still rely on proprietary models for the most capable steps today, open models are rapidly catching up. In April 2026, the open-weights model GLM-5.1 posted a score on Cybergym, a vulnerability analysis benchmark, surpassing closed alternatives such as Claude 4.6.[32] Attacks leveraging purely open-source models offer no provider-side mechanism for monitoring or restricting usage. Third, research on persistent prompt injection attacks suggests that agentic campaigns could eventually commandeer and weaponize other agents across the ecosystem, further distributing activity and obscuring its origin.[33]

[26] Zietlow, "How attackers stay invisible: rotation, encryption, low-volume activity, and identity blending."
[27] Anthropic, "Disrupting the first reported AI-orchestrated cyber espionage campaign."
[28] Folkerts et al., "Measuring AI Agents' Progress on Multi-Step Cyber Attack Scenarios."
[29] Jones et al., "Adversaries Can Misuse Combinations of Safe Models."
[30] A recent attack campaign illustrates such model disaggregation in action: open-source Deepseek was used for the comparatively simple task of reconnaissance while Claude managed capability intensive tasks, including vulnerability assessment and attack-time offensive tool execution. This evidence further suggests even when open source is used, attacks may still lean on closed model providers for capability augmentation, see Mike, "LLMs in the Kill Chain: Inside a Custom MCP Targeting FortiGate Devices Across Continents."
[31] Arghire, "Hackers Weaponize Claude Code in Mexican Government Cyberattack."
[32] Z.ai., "GLM-5.1: Towards Long-Horizon Tasks."

Disaggregation does not prevent individual providers from detecting misuse on their own platforms, but each provider may lack context to determine if requests are a component of a larger malicious operation. Disaggregated activity also confounds efforts to identify tactics, techniques, and procedures. As attackers learn to leverage more distributed operations, the detection advantage demonstrated in the GTG-1002 case, where Anthropic had direct visibility into the wider campaign, will erode.

*Defensive Implication: Disaggregated activity may mean no single point of observation, model provider, platform, or defender will have sufficient attack visibility for detection.*

### 3. The horizontal scale of offensive cyber agents will yield *everything, everywhere, all-at-once*-style campaigns

In the traditional cyber threat environment, sophisticated attacks were largely the exception rather than the rule, constrained by the time, attention, and effort of human operators. Conventional automation could scale low-complexity, high-volume attacks, such as phishing campaigns and credential stuffing,[34] but operations requiring judgment, adaptability, and sustained tradecraft could not be replicated at scale. Agentic systems, however, could soon shed these limits and be bottlenecked primarily by compute.[35] If attackers find the “price” of operating models is cheap, and success rates are high,[36] which early evidence suggests may be the case,[37][38] they may be significantly more likely to launch everything, everywhere, all-at-once campaigns, spraying sophisticated attacks across multiple targets and sectors simultaneously.

Early evidence supports this concern. In the recently documented GTG-1002 campaign, Anthropic noted the attack operated “simultaneously across multiple targets, with the AI maintaining separate operational contexts for each active campaign independently.”[39] Palo Alto Networks similarly observed that “Operator time is less of a constraint” as “AI-assisted workflows allow actors to run reconnaissance and initial access attempts across hundreds of targets in parallel.”[40]

---

[33] Yang et al., "Zombie Agents: Persistent Control of Self-Evolving LLM Agents via Self-Reinforcing Injections."
[34] Herley, "Security, Cybercrime, and Scale."
[35] de Moor, “The Chaos Phase: How AI is Transforming Cybersecurity Threats.”
[36] The relationship between inference costs and attack success is called Expected Cost per Success. Irregular, "When Success Rates Mislead: The Case for Expected Cost as a Metric in AI Evaluation."
[37] Nagli and Irregular, "AI Agents vs Humans: Who Wins at Web Hacking in 2026?"
[38] AI Security Institute, "Evidence for Inference Scaling in AI Cyber Tasks: Increased Evaluation Budgets Reveal Higher Success Rates."
[39] Anthropic, “Disrupting the first reported AI-orchestrated cyber espionage campaign.”
[40] Palo Alto Networks, “Global Incident Response Report 2026.”

This has significant implications for detection. Security teams already face significant alert volumes, but can generally distinguish between sophisticated attacks that require careful investigation and commodity threats that can be handled with fewer resources. If agents dramatically increase the amount of sophisticated attacks, this triage model faces significant strain. Further, existing cyber coordination and information-sharing institutions are often siloed within narrow industries, designated critical sectors, and national borders. Such architecture is not designed for attacks that are simultaneous, cross-sectoral, and systemic.

*Defensive Implication: High volume, horizontally scaled, sophisticated attacks could strain detection resources of security teams and existing threat-sharing institutions.*

# 2 | Detection-in-Depth

To meet these agent detection challenges, we propose a guiding strategy we call *detection-in-depth*. This approach rests on a central analytical premise: no single mechanism, process, or actor will be sufficient to detect autonomous cyber threats. Agents may adapt and vary their behavior to evade existing tools, disaggregate operations across multiple providers and environments, and execute sophisticated attacks at scale. Effective detection, therefore, requires not only new tools but a layered, complementary, and ecosystem-wide approach to generating, interpreting, and correlating signals of malicious activity.

Detection-in-depth's strategic logic builds on the cybersecurity principle of defense-in-depth, which holds that no single "silver bullet"[41] defense is sufficient and that effective security requires overlapping layers of control.[42] It extends that logic to the autonomous-cyber-threat era by making detection a central defensive priority. Recent research on defending against large numbers of intelligent, adaptable attackers supports this emphasis.[43] Against agents that can learn from failure, defensive layers should not be understood only as barriers meant to block attacks outright. Their value also lies in slowing attackers down and forcing repeated intrusion attempts at each stage. If detection capabilities are robust, each reattempt becomes a new opportunity to identify and expel the attacker before the operation succeeds. As research modeling shows, even modest improvements in detection can sharply reduce the likelihood of attacker success. Detection is therefore not an adjunct to layered defense but the mechanism that makes it viable against advanced cyber agents at scale.

Detection-in-depth translates this insight into a concrete strategic principle: that **effective agentic detection requires layered and complementary detection mechanisms that together provide visibility into malicious activity no single tool or actor could catch alone.** Some mechanisms must operate at the point of access, helping establish who or what is acting. Others must operate within networks, platforms, and model-provider environments, helping determine what that actor is doing and whether it is suspicious. Still others must operate across the wider ecosystem, making it possible to identify distributed campaigns, correlate dispersed signals, and surface patterns that would remain invisible from any single vantage point.

This strategy also implies **distributed responsibility**. Different actors are positioned to detect different aspects of agentic threats. Model and cloud providers are uniquely placed to observe activity upstream at the point of model or service invocation. Network operators remain best placed to detect threats as they materialize in network and endpoint activity. Ecosystem-level detection will depend on institutions and mechanisms that can combine signals across organizations, providers, and sectors. Detection-in-depth, therefore, requires not just better tools, but coordination across the actors who hold different pieces of the overall visibility picture.

[41] Industrial Control Systems Cyber Emergency Response Team, "Recommended Practice: Improving Industrial Control System Cybersecurity with Defense-in-Depth Strategies."
[42] Cyber Security Agency Of Singapore, "Guidelines on Securing AI Systems."
[43] Lohn, "Defending Against Intelligent Attackers at Large Scales."

We want to be clear about what detection-in-depth is and is not. It is not a single program or fixed architecture to be implemented wholesale. It is a strategic framing and a set of organizing principles for building detection capacity against autonomous cyber operations. As the threat evolves, so too should the mechanisms, institutions, and defensive practices associated with it.

## Detection-in-Depth Core Functions

The National Institute of Standards and Technology (NIST) cybersecurity framework defines detection as the "timely **discovery** and **analysis** of anomalies, indicators of compromise, and other potentially adverse events."[44] Detection-in-depth builds on this foundation, focusing on three essential functions layered detection mechanisms should serve to meet the agentic detection challenge:

- **Discovery, via signal generation:** Agentic detection will require the generation of novel signals that better define the shape of the threat environment, flag agentic threats in real time, and feed detection-relevant analysis. This data should aim to answer foundational questions of agent detection, including: What activity is agentic? What activity is human-directed? And are digital actors operating within their authorized boundaries?
- **Analysis**: Agent detection will require the analytical infrastructure, mechanisms, and processes needed to interpret these signals, confirm malicious activity, and produce actionable threat intelligence. This function should aim to answer foundational questions of agent detection, including: Is this agent's behavior malicious or authorized? What is the nature of this attack—what tactics, techniques, and procedures is it employing; what is the operational setup; what models and system components are employed?
- **Coordination**: Agentic detection will demand the processes and institutions needed to enable systemic action, including threat intelligence sharing, the pooling of disparate signals, and ecosystem-wide threat analysis. This function should aim to service detection-in-depth's systemic goals while aiming to answer foundational questions of agent detection, including: is the incident isolated or part of a coordinated campaign? Who is responsible, and under whose authority is this agent acting.

[44] NIST, "The NIST Cybersecurity Framework."

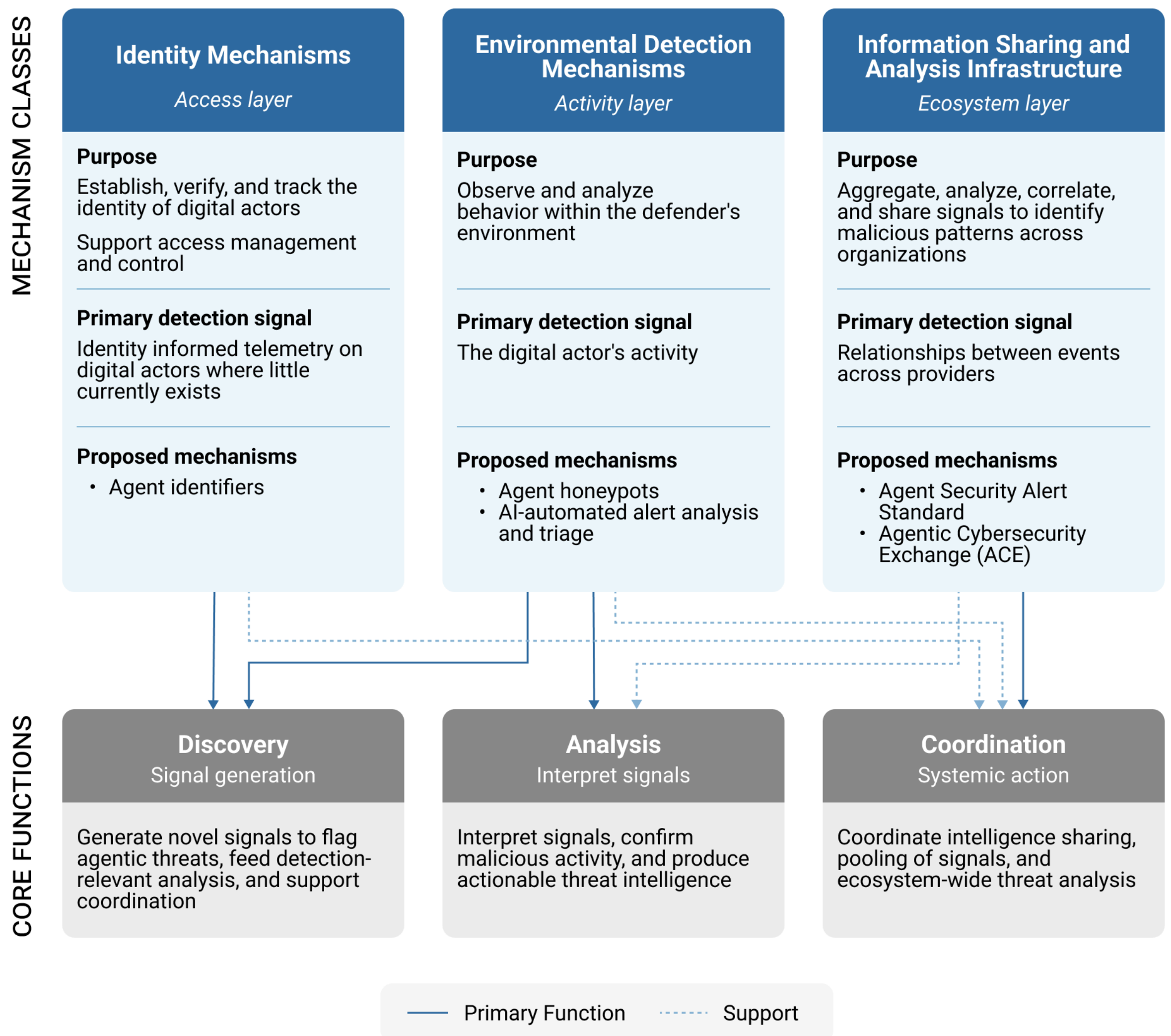

## Detection-in-Depth Mechanism Classes

There are many potential mechanisms that will be needed to put these functions into practice. To provide a starting point, we recommend three essential classes of mechanisms that together support a detection-in-depth strategy:

**Identity Mechanisms** operate primarily at the **access level**, before or at the point where an agent or actor enters a system or invokes a service. Their primary role within the detection-in-depth approach is signal generation. These mechanisms work by establishing and verifying identity, checking whether that identity is authorized for the action being attempted, and logging that actor's activity over time. The core detection question at this layer is*:* Who is this digital actor, what are their intentions, and can their activity be tracked? Detection signal comes from generating detection-relevant telemetry on digital actors where little currently exists, and from anomalies such as absent or spoofed credentials or agents exceeding their authorized scope. Agent Identifiers are the primary example mechanism we explore later in this report (see "Section 3: Agent Identity for Critical Infrastructure"); however, these mechanisms could also include any technology, standards, or processes that better define and trace who is doing what online. When agents can be identified and traced, a range of detection possibilities follow, from attributing operator control and benchmarking normal behavior to aid anomaly detection, to tracing interaction graphs that reveal compromises, forensics, and campaign scope.

Identity infrastructure is nonetheless insufficient for detection on its own. It can be spoofed or circumvented, and agents operating outside formal infrastructures will be invisible. Where identity mechanisms fall short, other mechanisms, such as environment-level detection, are required.

**Environmental Detection Mechanisms** operate primarily at the **activity level,** once an actor is already present in a target network, cloud platform, or model provider's infrastructure, and aim to fill the discovery via signal generation and analysis functions of detection-in-depth. The core detection question at this layer is: What is this actor doing, and is it suspicious? These mechanisms work by observing behavior within the defender's environment. The detection signal comes from what the actor does, not who it claims to be.

This class of mechanisms includes a range of approaches. Behavioral monitoring and endpoint detection and response (EDR) tools flag deviations from established baselines. Agent-specific anomaly detection targets patterns distinctive to agentic activity, such as machine-speed command sequences. Later in this report, we focus on agent honeypots (decoy systems designed to lure autonomous attackers) as a promising example of an agent-specific environmental detection mechanism (see "Section 4: Agent Honeypots").

Generating more signals, however, is only valuable if defenders can process them. Automated alert analysis and triage uses AI-assisted methods to filter, prioritize, and assess the growing volume of detection signals, helping defenders maintain analytical capacity as the volume and sophistication of agentic attacks scale (see "Section 5: Automated Alert Analysis and Triage"). These mechanisms are the workhorse of detection-in-depth, translating signals into actionable alerts as attacks unfold.

Identity context makes environment-level detection more powerful: a behavioral anomaly that might otherwise be noise becomes a high-confidence signal when linked to a known agent, operator, or campaign. But these mechanisms are limited in scope. They detect activity within a single organization's perimeter and cannot surface parallel attacks on others.[45] Disaggregated attacks may evade any single organization's detection entirely. This gap motivates the need for our final class of mechanisms.

---

[45] Large-scale honeynet deployments, such as AWS MadPot, can provide broader visibility by operating across extensive infrastructure, see Ryland, "How AWS threat intelligence deters threat actors."

**Information Sharing and Analysis Infrastructure** operates at the **ecosystem level** and aims to fill the coordination function and supplement the analysis function of detection-in-depth. The core detection question at this layer is: What does this activity look like when combined with what others are seeing? These mechanisms work by aggregating, analyzing, correlating, and sharing signals across organizations, platforms, and infrastructure providers to identify malicious patterns that no single actor can detect alone. The detection signal comes from relationships between dispersed events—activity that appears benign or ambiguous in isolation but becomes suspicious when viewed across multiple environments, targets, or service providers. Because agentic attacks may parallelize across many targets simultaneously and distribute activity across models, tools, and vendors, ecosystem-level visibility is necessary to identify coordinated campaigns and emerging threat patterns.

One promising approach is supporting centralized information sharing and analysis institutions which can aggregate telemetry and threat intelligence across actors to detect and share cross-platform threats. We propose the creation of an Agentic Cybersecurity Exchange (ACE), an industry led institution focused on detecting and disrupting agentic cyberthreats, as one such mechanism (see "Section 7: Agentic Cybersecurity Exchange (ACE)").

A complementary approach is to invest policy effort into *enablers* of information sharing, mechanisms that reduce sharing frictions and allow information to diffuse to those outside the operational scope of centralized institutions. Example enablers include shared threat alert standards, threat intelligence communication protocols, permissive legal frameworks, and common definitions. Later in this report, we propose an Agentic Security Alert Standard as one tractable example (see "Section 6: Agentic Security Alert Standard").

While ecosystem-level coordination expands detection beyond individual organizational perimeters, it is also constrained by participation incentives, privacy and legal barriers, and potential gaps in coverage. For this reason, information-sharing mechanisms should reinforce rather than replace identity- and environment-level detection mechanisms, ensuring that detection remains resilient even when coordination is imperfect.

---

The sections that follow give practical form to this approach, covering five proposed mechanisms: agent identity for critical infrastructure, agent honeypots, automated alert analysis and triage, agent security alert standards, and an Agentic Cybersecurity Exchange (ACE). Together, these proposed mechanisms are intended to seed each recommended mechanism class of detection-in-depth.

# 3 | Agent Identity for Critical Infrastructure

*Stakeholders: Critical infrastructure operators, governments, standards bodies, agent service providers. For implementation, see recommendations.*

Today, AI agents operate largely without persistent, verifiable digital identifiers. This means that agentic activity across networks, platforms, and services is frequently indistinguishable from anonymous software traffic. When systems of agent identity do exist, they are often transient,[46] platform-specific,[47] or loosely coupled to the entity ultimately responsible for the agent's behavior. This identity vacuum imposes a fundamental constraint on defensive analysis. Without tools to identify and track agents and their authorized capabilities, defenders can only rely on indirect or probabilistic signals to detect malicious activity.

Agent identifiers (Agent IDs) offer one way to address this problem and to support the signal-generation function of detection-in-depth. An Agent ID is a persistent identity token that an agent shares when interacting with users, platforms, or services. That token carries information that both identifies and describes the agent. Depending on the design, this may include a unique identifier for the agent instance, information about its sponsor or operator, its developer or hosting provider, and other descriptive metadata (see What is an Agent ID? for more details).

The core security value of Agent IDs is not that they would directly reveal malicious activity. Rather, their main value is as a telemetry layer that supports the signal generation function of detection-in-depth. Agent IDs could make a meaningful share of agent activity visible to defenders, where little or no agent-specific telemetry currently exists. That telemetry could then be combined with other indicators to support triage, anomaly detection, forensic reconstruction, and coordinated response. In other words, Agent IDs should be understood not as a primary prevention mechanism, but as an identity layer that could improve detection and incident response in environments where agentic activity may otherwise be opaque.

As with any detection mechanism, Agent IDs are not a panacea. Their primary detection benefits accrue to just potential targets, not agent providers, while the signals they generate are unlikely to identify malicious activity in isolation. An Agent ID alone cannot distinguish a compromised agent from a legitimate one, nor can it reveal whether an agent's declared metadata accurately reflects its actual behavior. Rather, Agent ID signals must be combined with broader behavioral monitoring, threat intelligence, and anomaly detection to become actionable. Importantly, **agents operated on attacker-hosted models will be unidentified in most cases.** Agent IDs will likely be concentrated first among model provider-hosted and enterprise-deployed agents. Even so, identifying legitimate model provider-hosted and enterprise-hosted agents would still improve defensive triage by helping defenders distinguish known agent traffic from unattributed automation, and by supporting detection of provider-hosted agent misuse, agent hijacking, and other compromises involving credentialed agents.

---

[46] OAuth-based systems commonly rely on short-lived access tokens.

[47] Application and workload identities are typically implemented through platform-specific constructs such as service accounts or IAM roles.

These benefits can be reaped only if a plurality of agents bear IDs. Success, however, still depends on a meaningful level of adoption. If only a negligible fraction of agents interacting with an organization bear IDs, the resulting telemetry will be too sparse to materially improve detection. Achieving useful scale requires two forms of adoption at once: agent providers must begin issuing IDs, and the external services those agents interact with must be capable of reading and acting on them. Ecosystem-wide ID adoption should be the long-run goal. Recognizing adoption challenges, however, we recommend critical infrastructure as an adoption starting point. Critical infrastructure operators should begin requiring agent IDs for all external agents they interact with, implementing mechanisms to recognize those IDs, and ensuring detection systems can make use of agent identity information in their analysis. As noted in the recommendations below, this process will require critical infrastructure operators to partner with governments, standards bodies, and agent service providers to ensure standards are consistent.

## Why Critical Infrastructure?

There are three reasons critical infrastructure is an optimal starting point:

**Emerging agent deployment in some critical infrastructure sectors:**[48] In healthcare, agents may soon be deployed to handle complex inter-organizational workflows such as claims management,[49] and manage consumer-health system interactions. In finance, firms are already piloting agentic payment transactions,[50] while other applications like bank-to-bank payments,[51] automated regulatory oversight, and customer service[52] may emerge soon. Firms' systems can also be expected to interact with agents that consumers deploy to manage personal finances. Finally, in government services, some nations are already testing agent-managed constituent

[48] Agent adoption is likely to vary significantly across critical infrastructure sectors, with transaction-heavy and heavily digitized sectors, such as healthcare and finance, seeing earlier uptake than sectors reliant on physical processes and legacy control systems.
[49] Martin et al., "What are AI agents, and what can they do for healthcare?"
[50] Paz.ai, "Visa, Mastercard, and PayPal's Q4 2025 Moves Into Agentic Commerce: What Retailers Must Know."
[51] Fireblocks Staff, "The Infrastructure Layer for AI Agents in Institutional Finance."
[52] Quick et al., "The Next Wave Arrives: Agentic AI in Financial Services."

interactions[53]—agents that may themselves soon interact with agents deployed by citizens to help navigate government processes.

**Elevated and consequential cyber risk:** Critical infrastructure is a priority target for adversaries. In 2025, for instance, more than half of ransomware attacks were directed at critical infrastructure sectors.[54] What's more, attacks on these sectors are most likely to yield catastrophic effects. Illustratively, a 2026 study found that ransomware attacks on hospitals were associated with a massive 34–38% increase in in-hospital mortality.[55]

**Structural capacity to drive broader adoption:** Adoption will ultimately be demand-driven. While it is agent providers who must actually issue IDs, that will only be done persistently at scale if a critical mass of receiving services (the platforms, APIs, and systems that agents interact with) start to actually demand, accept, and act on IDs. Critical infrastructure sectors are well-positioned to generate that pressure: they are often sufficiently large and interconnected enough to create spillover pressure on adjacent industries and service providers. Policy action to push Agent ID adoption within critical infrastructure sectors is also uniquely plausible. Today, many sectors are already governed by existing regulatory frameworks and institutions. This policy infrastructure could provide a foundation for any new regulations or incentives that may be required to push adoption forward.

The point is not that critical infrastructure is the only context in which Agent IDs would be useful. Rather, it is that these sectors combine high exposure, high consequence, and plausible pathways to adoption. If policymakers want to test whether Agent IDs can generate meaningful detection value, critical infrastructure is a sensible place to start. To better understand the value of such action, below we discuss what agent IDs are and their potential detection benefits. Next, we discuss the adoption requirements for critical infrastructure providers to reap these benefits.

## What is an Agent ID?

While the shared goal of any agent ID schemes is to identify agents, technical implementation details can significantly impact potential use cases. Differences in proposed implementations can change when, how, and to whom an agent ID is communicated, the ID's technical form, the means of verifying the agent ID's data, and other essential details. **This report does not seek to discuss specific technical agent ID implementations; its focus is on how agent IDs can act as a detection mechanism for critical infrastructure operators**. To ground understanding, however, a vignette of one real-world standard being used in practice is illustrated in BOX 1 below.

[53] Collins, "The new era of public sector service is about building trust through agentic AI."
[54] Kela Cyber, "Escalating Ransomware Threats to National Security."
[55] Neprash et al., "Hacked to Pieces? The Effects of Ransomware Attacks on Hospitals and Patients."

That said, regardless of how an agent ID scheme is implemented, any scheme intended to support detection should include the following minimum elements:

**First, an agent ID must include some form of persistent, ideally cryptographically verifiable, identifier attached to each agent instance.** This identifier will often be represented as a string of alphanumeric characters, much like a driver's license number.[56] This identifier functions as a stable digital credential that the agent will actively send to any entity or user during interactions with that agent in the form of a digital token. This ideally allows users to both understand that they are interacting with an agent, and distinguish the specific agent they are interacting with.[57] Contrasted with perhaps more familiar session tokens, which are temporary identifiers tied to a single digital interaction, or platform-specific identifiers that only work within one system, this identifier is intended to persist across time and across the different environments with which the agent interacts. This identifier is the essential anchor for any agent ID scheme. For defenders, only if an agent is paired with an identifier can they then track and monitor their interactions with that agent over time.

**Second, an Agent ID scheme should communicate structured metadata describing each identified agent.** In many implementations, both this data and its associated identifier will be packaged in a digital token that is shared with users and entities during interactions. This data provides essential context about that agent and its expected behavior, such as the agent's provenance, the underlying models it uses, declared permissions, and declared capabilities (e.g., tool-use capabilities, knowledge bases). Importantly, this data can also include additional identifiers to flag those responsible for the agent. This potentially can include its developers, the hosting provider, and the user or entity authorizing its actions. Finally, this metadata can help establish a delegation chain, linking users to any agents with parent agents and any child agents those parents deploy. For defenders, this metadata represents a potential wealth of telemetry that could be essential to detection capabilities described below.

**Third, an Agent ID scheme must include registry infrastructure to manage the lifecycle of Agent IDs, including issuance, updates, revocation, and validation.** Only with a trusted registry can users validate the accuracy of claims made in an Agent ID. For instance, if an agent's ID declares it is authorized to handle transactions by a user's bank, a registry can verify whether this claim is indeed true to ensure the user's financial safety. For defenders, only with the trust registries provide can the telemetry agent IDs be trusted enough to meaningfully inform detection efforts.

---

[56] For instance, under Microsoft EntraID's implementation of Agent ID's an agent's identifier can look like: aaaaaaaa-1111-2222-3333-bbbbbbbbbb

[57] Note that many current implementations have yet to meet this ideal. Some, such as cloudflare's Web Bot auth protocol, do not distinguish each discrete agent, but rather whether an identified agent represents a given organization.

## Illustrating Agent IDs in Practice: A Healthcare Claims Agent Under OIDC-A

Consider Tokenridge General, a forward-looking hospital that deploys an instance of an OpenLobster agent to handle insurance claims submission. Under OpenID Connect for Agents (OIDC-A), an Agent ID protocol, here is how that agent might establish and communicate its identity.

**Registration.** Upon deployment, Tokenridge General registers its agent instance with an OIDC-A compliant identity provider, which issues the agent a cryptographically signed identity token containing a unique agent identifier: Agent85CTQ. That Identifier is packaged with standard metadata describing the ID's status: active; its authorizing organization: Tokenridge General; its underlying model: OpenModel-7B; its scope of authorized actions: submitting claims; and an authorized staff credential delegating authority to the agent to act on their behalf: Angela Smith.

**Interaction.** The agent contacts the claims API of LLMutual, an insurer, and presents its signed token. LLMutual has deployed an OIDC-A compliant verification server that checks the token's cryptographic signature against a trusted registry—the authoritative record of issued agent identifiers and their current status. The registry confirms that Agent85CTQ is active and was legitimately issued, allowing the server to then read and act on the token's metadata, checking that the agent is operating within its declared scope. Once validated, the agent is cleared to proceed.

**Defensive Illustration.** In the background, the agent's identifier allows the LLMutual's security team to carefully track and log its interactions. While submitting the claim, the agent begins to simultaneously submit requests for the patient's past insurance records—actions outside its declared scope. Because the token describes the agent's authorized actions, the security team is able to immediately detect the anomaly and block the agent. They then inform the Healthcare-ISAC that any agent bearing the identifier Agent85CTQ is potentially compromised. Because Tokenridge General is listed as the authorizing organization, the LLMutual team is also able to contact them directly. Reviewing their logs, Tokenridge's security department finds the agent had interacted with a malicious prompt and was subsequently hijacked. The team terminates the agent and conducts a security review.

## Detection Value of Agent IDs

Agent IDs can service agentic detection in a variety of ways, including:

**Providing Detection-Relevant Network Telemetry:** The most consequential detection function of agent identity systems is producing telemetry about agent activity that defenders would otherwise lack. Without persistent identifiers, incoming requests from AI agents are largely indistinguishable from ordinary traffic, detectable only through imprecise behavioral inference. With a structured identity framework, however, a potentially meaningful share of that traffic could become legible: tied to specific agent instances, defined sponsors, and delegation chains. These data will equip defenders to establish baseline flows of agent-based activity and monitor traffic from known agent providers.

This telemetry data will be especially important for detecting attacks that utilize ID-bearing proprietary agents. Misuse of proprietary agents is already common, and agent hijacking has appeared in recent attacks, including the S1ngularity campaign, in which benign agents were manipulated through malicious prompts to facilitate credential theft.[58] These risks may grow as agents become more exposed to prompt injection, self-replicating prompts, and other ecosystem-level threats.[59] In these cases, Agent IDs could help defenders monitor agents to detect such events, identify which known agents were involved, distinguish misuse from unrelated automation, and track potentially compromised agents across incidents.

Identity systems aid detection not only through the data they generate, but through the gaps they reveal. The absence of an Agent ID or an invalid identity becomes meaningful. Missing identifiers, broken credential chains, unverifiable provenance, and repeated re-registration attempts can all serve as signals that can help defenders segment and triage traffic for different levels of analysis and scrutiny. This defensive triage can be used to better target the environment-level detection mechanisms needed to distinguish any hidden or malicious agents from benign traffic. It can also prompt stricter logging and tighter rate limits. A further signal lies in the ratio between identified and unidentified traffic. If either side of that ratio changes dramatically, that could signal machine-scale malicious activity.

**Behavioral Monitoring:** Persistent identifiers make it possible to perform identity-conditioned anomaly detection. Using Agent IDs, receiving services can build agent-specific behavioral baselines that can be used to flag anomalies such as atypical resource access, irregular request timing patterns, atypical tool use, or anomalous request volume. If Agent ID metadata includes declared permissions or capability bounds, unauthorized behavior becomes a distinct detection signal: any agent accessing systems outside its designated scope, requesting unnecessary privileges, or operating at a scale inconsistent with its declared function can be automatically flagged.[60] Already, such identity-conditioned anomaly detection functions are being developed into product proofs of concept. Microsoft EntraID, for instance, allows administrators to monitor identified agents and flag suspicious events, including failed access attempts, sign-in spikes, and unfamiliar resource access. It also allows malicious agents to be tagged and tracked as "confirmed compromised."[61]

[58] Winterford, "The s1ngularity attack: When attackers prompt your AI agents to do their bidding."
[59] Cohen et al., "Here Comes The AI Worm: Unleashing Zero-click Worms that Target GenAI-Powered Applications."

**Forensic Reconstruction:** Persistent identifiers may also support forensic reconstruction after an incident. Using identity-enriched logs, defenders could reconstruct an agent's provenance, the external services and resources it accessed, the tools and subagents it employed, and the sequence of actions it took. This could support faster scoping of blast radius, clearer reconstruction of attacker methods and objectives, and greater visibility into delegation relationships between parent and child agents.

**Ecosystem and Campaign Correlation:** If agents can be tied to stable Agent IDs, detection analysis might be extensible across organizational boundaries, helping bootstrap the coordination when responding to any attacks involving identified agents. If bearing an ID, a malicious agent's logged activity in one environment could be correlated to logs held by other organizations, platforms, or service providers. This could enable ecosystem-wide detection, coordinated incident response, and improve general situational awareness about attack methods, campaigns, and malicious actors.[62] Further, agent metadata, such as controller linkages, provenance data, and delegation chains, might allow analysts to better connect the dots between seemingly separate incidents to help identify subtle attack signatures and coordinated campaigns.

**Limitations:** These detection possibilities have limits. Agent identity will likely be substantially weaker for browser-mediated or computer-use workflows. Where an agent interacts with services through standard APIs, identity-bearing tokens can be attached to each request and validated by the receiving service. Where an agent acts through an ordinary browser session, however, activity may be indistinguishable from a human user, with no clean mechanism for presenting or validating agent identity. This means identity-based detection is most effective against agents operating through API channels and least effective against agents that deliberately route through browser-based interfaces.

---

[60] Agent identifiers may also strengthen provenance-based intrusion detection (PIDs), a class of detection methods that continuously track objects within a digital system (processes, files, network connections) and their interactions over time, constructing "provenance graphs" that model normal system behavior. These graphs allow defenders to identify malicious chains of activity that would be invisible when examining individual events in isolation or over short durations. Through agent IDs, agents can become a legible network object, potentially bolstering provenance-based detection. This value-add is worth investigation and consideration as provenance-based methods have demonstrated particular effectiveness in detecting Advanced Persistent Threats (APTs) (See Cheng et al., "KAIROS: Practical Intrusion Detection and Investigation using Whole-system Provenance.").

[61] Microsoft Learn, "ID Protection for agents (Preview)."

[62] Chan et al., "Infrastructure for AI Agents."

Further, use cases, such as campaign correlation and behavioral monitoring, are likely to be most effective against attacks that use persistent agent identities. To avoid detection, sophisticated attackers may rotate agent identities to limit the correlation window and the ability of defenders to benchmark agent behavior, much as they already rotate network indicators. Depending on the security of the identity scheme, there is also an ever-present risk that actors may figure out how to compromise, forge, and steal identity tokens. Finally, agents operating within legitimately granted permissions can escalate privileges, disable security defenses, and exfiltrate data in ways that may appear routine in audit logs until forensic reconstruction pieces together the full sequence of actions.[63] For these reasons, any critical infrastructure implementation strategy must therefore follow detection-in-depth's recommended strategy: pairing identity infrastructure with complementary layers of detection mechanisms.

## Design and Implementation Considerations

To take advantage of these detection benefits, critical infrastructure providers must mandate that any agent interacting with their systems bear an ID. While perfect compliance should not be expected, benefits will accrue even if only a plurality of agents carry IDs.

Achieving that scale in critical infrastructure will likely require policy intervention rather than voluntary adoption. Given the unique systemic risks posed by attacks on critical infrastructure and the likely slow pace of organic uptake, policymakers should consider mandating Agent ID adoption in target sectors rather than waiting for market pressure to develop.

One option is a regulatory mandate. While any regulation faces an uphill battle, it is likely in many cases that an agent ID requirement could easily build on existing statutes and regulations in sectors already subject to security, privacy, reporting, and auditing rules.[64][65] A second option, implementable alongside or in lieu of mandates, is financial incentives: direct payments, grants, or tax credits targeted at operators who adopt compliant ID systems. This matters because Agent ID adoption will require potentially significant investment, not only in the software needed to receive and interpret IDs, but in the broader infrastructure required to realize their detection potential. This may include logging systems, Agent ID-informed access controls, and agent-specific threat intelligence and information-sharing platforms. For the many under-resourced operators across critical infrastructure sectors, financial incentives may be essential.

---

[63] Irregular, "Emergent Cyber Behavior: When AI Agents Become Offensive Threat Actors."

[64] For instance, in the United States healthcare IT systems must already "Assign a unique name and/or number for identifying and tracking user identity." This regulation need only be extended slightly to require agent IDs distinct from already required human user IDs. See eCFR, "§ 164.312 Technical safeguards."

[65] Detection-relevant implementation may be especially tractable in the financial services sector where anti-money laundering and know your customer rules are well established and could be extended to mandate verification of the link between an agent and its responsible human operator.

Regardless of the specific adoption incentives provided, however, policy success demands a system that considers the following:

## Appropriate Policy Scope

Critical infrastructure is often a broad and ill-defined category.[66] To implement any policy effectively, decision-makers must specify precisely where adoption should be prioritized. The financial and healthcare sectors are logical starting points on both counts. Each is a major consumer of information technology and already relies on complex, layered ecosystems of consumer-facing applications, third-party platforms, APIs, and data services—precisely the environments where AI agents are most likely to be deployed at scale in the near term. Both sectors are also systemically important: failures or compromises in either can cascade across the broader economy, raising the stakes for undetected malicious agent activity. Critically, these sectors are often subject to robust regulatory frameworks, giving regulators existing institutional mechanisms through which to introduce Agent ID requirements without building enforcement infrastructure from scratch.

## Standards and Interoperability

To scale adoption, firms and policymakers must standardize Agent ID formats and protocols. The technical building blocks already exist (PKI, SPIFFE, OpenID Connect),[67] yet dozens of incompatible frameworks are emerging in parallel, including OpenID for Agents (OIDC-A)[68] and MCP Identity (MCP-I),[69] with no clear frontrunner. Without a common standard, infrastructure operators would face an unappealing choice: implement parallel interpretation layers for every competing format, or accept that some agent identity data will simply be unreadable. Either outcome would erode the practical value of agent IDs and risk discouraging the very adoption that makes the detection value of IDs viable in the first place.

Crucially, this standard-setting effort need not wait on model providers that have thus far failed to converge on any single system. Critical infrastructure operators, by virtue of their economic scale and the sensitivity of their environments, are well-positioned to drive this process instead. Working alongside trusted bodies, such as the National Institute of Standards and Technology, Internet

[66] Mittelsteadt, "Critical Risks: Rethinking Critical Infrastructure Policy for Targeted AI Regulation."
[67] PKI (Public Key Infrastructure) is the system used to issue and verify digital certificates, i.e., the same technology that secures website connections. SPIFFE (Secure Production Identity Framework for Everyone) is a framework for issuing short-lived identity credentials to software workloads. OpenID Connect is a widely adopted protocol for verifying identity across different services, used by many consumer and enterprise login systems. All three are mature, widely deployed technologies that could, in principle, be extended to support agent authentication.
[68] Nagabhushanaradhya, "OpenID Connect for Agents (OIDC-A) 1.0: A Standard Extension for LLM-Based Agent Identity and Authorization."
[69] Model Context Protocol Identity, "Model Context Protocol – Identity (MCP-I) Specification."

Engineering Task Force, and the OpenID Foundation, they could build consensus around a common baseline standard applicable to any agent interacting with critical sectors.

## Detection-Relevant Telemetry

Policymakers must ensure any selected standard provides the rich data needed for detection. As noted, Agent IDs must include both an identifier and attached metadata. The identifier alone is the single most important data element for detection. Even a standard that only defines an identifier, basic provenance metadata, and status metadata (indicating whether that identifier is active or revoked) would unlock significant defensive capabilities. These include establishment of baseline flows of normal agent activity, agent-specific monitoring and logging, identification of systems with which malicious agents have interacted, agent  behavioral baselining, identification and monitoring of compromised agents, and the ability to correlate agent activity across organizations.

That said, sufficiently detailed metadata can both ease identity-informed detection and unlock further possibilities. Full detection success, therefore, requires the following detection-relevant metadata categories:

- **Accountability-related metadata**: Describes the responsible humans behind an agent, including developers, deployers, and authorizing users. This can help defenders identify which parties to contact when an agent is found to be compromised. It can also help defenders flag agents originating from actors known to be compromised, malicious, or to have lax security practices for greater scrutiny.
- **Delegation chain related metadata:** Describes parent-child relationships between operators, agents, and subagents.[70] Combined with accountability metadata, this helps defenders connect the dots between agent activity and broader potential attack campaigns.
- **Behavior-related metadata:** Describes the expected behavior of a given agent, including model information, declared capabilities, permissions, and tool-use scope. This would further enhance the behavioral monitoring functions of Agent IDs.

## Authentication

The detection value of Agent IDs depends on whether the identity data they carry can be trusted. A system in which identifiers can be spoofed, discarded, or paired with fabricated metadata offers little defensive value and may actively mislead defenders. Agent IDs, therefore, require mechanisms to authenticate both the identifier itself and at least some of the claims attached to it.

[70] While the technical specifics of agent ID delegation chains are out of this report's scope, it is important to note this class of metadata will demand airtight processes to a) bind an agent to its responsible operator, b) link a subagent's ID to a parent agent's ID, and c) enable receiving services to validate that delegation chains are unbroken and subagent permissions have not been expanded or altered. Such processes are an area of active research and are already features of existing standards such as OIDC-A.

In closed or vertically integrated environments, this problem is partly solved. Enterprise systems such as Microsoft Entra can authenticate agents, manage credential lifecycles, and support revocation and monitoring within a single administrative domain.[71] Across organizational boundaries, however, authentication is much weaker, and the fragmentation in identity formats discussed above makes external signals harder to interpret and trust.

More importantly, current proposals are better at verifying who issued or operates an agent[72] than the fuller set of properties defenders need to know about what the agent is—such as who is accountable for it, what permissions it is supposed to have, and whether it belongs to a delegation chain. Authentication proposals also do not currently extend to the broader runtime environment: what code wraps the model call, what tools the agent invokes, or how it behaves between interactions. Provider-level attestation is technically feasible in controlled cloud deployments and is the subject of active research,[73] but it is not yet widely deployed or standardized. More granular runtime attestation remains further from implementation because, unlike static software that can be verified once at launch, an active agent's behavior changes continuously as it calls tools, spawns sub-agents, and responds to new inputs.

Given present authentication challenges, policymakers should treat authentication as a staged requirement that must evolve over time. Early standards may only be able to strongly verify a limited set of properties, such as issuer identity, operator identity, and credential status, but they should be designed to expand over time toward richer accountability, delegation, and behavioral claims as the technical and institutional foundations mature.

## Registry Governance

Identity registries provide the trust infrastructure needed to support Agent ID functionality and authentication. Registries must ensure identifier uniqueness, prevent collisions or squatting, record lifecycle status, and support revocation. These functions are critical for detection because defenders must be able to determine whether an identifier is valid, who issued it, and whether it has been suspended or associated with compromise.

To perform these functions, registries must do more than administer names. They must set admission criteria, allocate identifiers, record status changes, support interoperability, and maintain auditable, tamper-resistant records and reliable lifecycle management so that identity signals remain trustworthy and operationally useful.

[71] Microsoft Entra illustrates a vertically integrated model in which certificate authorities, token services, conditional access, telemetry, and revocation are bound tightly into a single platform stack, see Microsoft, "Microsoft Entra: Identity and Network Access Solutions."
[72] For instance, Cloudflare's Web Bot Auth protocol allows agent *operators* to cryptographically sign HTTP requests so that receiving services can verify the operator responsible for an identified agent. Such operator-level authentication is useful, yet incomplete. See Meunier and Galicier, "Forget IPs: Using Cryptography to Verify Bot and Agent Traffic."
[73] Bodea et al., "Trusted AI Agents in the Cloud."

Given the sensitivity of critical infrastructure security, policymakers must carefully consider who governs registries and their security requirements. Potential models include centralized registries operated by major agent platforms, a federated consortia of registries , or decentralized identity systems. Regardless of institutional form, whoever operates registries must be trusted. Their decisions will shape reliability, neutrality, and resilience. Further, registries must be secure. The greater the degree of centralization, the greater the risk of attacks. As agent IDs are adopted, registries can be expected to emerge. However, policymakers must consider whether the sensitivity of critical infrastructure demands a bespoke treatment, such as a sectoral or government-run registry, for any agents intended to interact with critical infrastructure.

## Privacy, Legal, and Commercial Constraints

Agent IDs and their metadata may also face commercial, legal, and security constraints. Delegation chains and capability metadata could expose sensitive commercial information about product design, business processes, or organizational relationships. These concerns may reduce adoption or narrow the kinds of metadata organizations are willing to disclose.

Agent IDs and their detection use cases may also be subject to any law regulating personally identifiable information (PII) or other regulated data classes. For instance, retaining detailed identity metadata for forensic purposes may be in direct tension with data minimization principles under frameworks such as GDPR. [74]

Mitigations for these privacy concerns exist, including privacy-preserving verification techniques, purpose-bound retention policies, and correlation methods that enable security analysis without exposing which organizations are involved.[75] These are technically available but add complexity and are not yet widely deployed. If required, policy must consider accounting for the compliance burden these technologies will place on private sector adopters. If that burden is high, financial incentives should be considered. Further, policymakers must assess whether targeted legal clarifications or exceptions are needed to allow Agent ID adoption to proceed without creating unacceptable legal exposure for critical infrastructure.

---

[74] Regulation (EU) 2016/679, General Data Protection Regulation, Article 5(1)(c) (data minimization)
[75] For example, zero-knowledge proofs are a cryptographic technique that allows one party to prove a claim (e.g., "this agent is authorized for financial transactions") without revealing the underlying data (e.g., which organization authorized it or for what specific purpose). Pseudonymous correlation protocols would allow a central analysis body to detect that the same agent identity appears across multiple incident reports without learning which organizations filed those reports, see McMullen, "AI Agents Need Identity and Zero-Knowledge Proofs Are the Solution."

## Recommendations

1. **Governments, standards bodies, agent service providers, and critical infrastructure owners and operators should prioritize convergence on interoperable agent ID standards before proprietary fragmentation hardens**.

Initiatives such as OIDC-A, MCP-Identity, and Web Bot Auth are establishing agent ID technical foundations, but without coordinated adoption, they risk fragmenting the ecosystem into incompatible systems unable to achieve detection-relevant scale. Frontier model providers and major cloud platforms are already implementing forms of agent identification (e.g., Microsoft Entra Agent ID, Cloudflare Web Bot Auth, Okta), making them natural first adopters. However, their current implementations are not fully interoperable. Governments and critical infrastructure owners and operators should actively support and push for standardization through bodies like the Internet Engineering Task Force and the OpenID Foundation. They must also press major agent service providers for interoperability commitments.

2. **Governments should require and/or incentivize Agent ID adoption among Critical Infrastructure owners and operators.**

Critical infrastructure sectors, such as healthcare and finance, are the most plausible places to begin, as they combine high exposure to agentic AI, elevated cyber risk, and existing regulatory capacity. Governments should consider explicit regulatory requirements and build on preexisting auditing, reporting, and system security requirements when applicable. They should also consider targeted financial incentives to accelerate adoption in these environments and to support under-resourced critical infrastructure owners and operators with potentially costly technical transitions.

3. **Governments should prepare institutional infrastructure for cross-organizational agent identity correlation.**

The campaign correlation and forensic capabilities described in this section require the ability to share and correlate agent identity metadata across organizational boundaries. This may not emerge from market forces alone. In the near term, governments should expand the mandates and technical capacity of existing threat-sharing bodies (including critical infrastructure ISACs, national cyber agencies and CERTs, and the Agentic Cybersecurity Exchange (ACE) described in Section 7 to incorporate agent identity data into their correlation workflows. Governments should also work toward international coordination frameworks to prevent fragmentation across jurisdictions. Governments should ensure cyber information sharing statutes, privacy regulations, and digital trade agreements are appropriately permissive to accommodate these functions.

4. **Governments and research funders should invest in privacy-preserving technology and more robust runtime attestation to support agent identity infrastructure.**

Privacy-preserving correlation technologies (such as zero-knowledge proofs and secure multi-party computation applied to agent identity data) would allow organizations to detect shared threats without exposing sensitive operational data to each other, directly addressing the privacy barrier to cross-organizational correlation. Second, more robust runtime attestation for AI agents would allow receiving services to verify not just who operates an agent but what model and code it is running. As discussed in the subsection on "Design and Implementation Considerations," provider-level authentication is deployable today, but granular attestation for agents that dynamically invoke tools and spawn sub-agents remains an open research problem. Progress on both fronts would expand what agent identity can achieve, particularly for forensic and campaign correlation capabilities.

# 4 | AI-Automated Alert Analysis and Triage

*Stakeholders: AI Companies, defenders, security researchers. For implementation, see* *Recommendations.*

Automated alert analysis and triage uses AI to assist security analysts in filtering, prioritizing, and interpreting the growing volume of detection signals expected from autonomous cyber operations. Alert analysis and triage are already critical bottlenecks in cyber defense, and agent attackers that scale sophisticated intrusions would strain capacity further. Automated alert analysis and triage contributes to detection-in-depth by maintaining defender capacity as attack volume and complexity grow—ensuring that detection signals generated by other mechanisms in this framework, such as agent identifiers and honeypots, can actually be acted on.

## What is AI-Automated Alert Analysis and Triage?

### Conventional Alert Analysis and Triage

When a detection system flags suspicious activity, security analysts must determine whether it warrants investigation, what happened, and how to respond. This process typically follows a pipeline: alerts are first triaged to filter false positives from genuine threats, then analyzed for context and severity, and where warranted, escalated into full investigations.[76] Delays at any stage extend attacker dwell time—the window between initial compromise and detection—giving adversaries more opportunity to reach their objectives.

Conventional automation systems can flag known threats and execute predefined responses, such as isolating a machine that matches a known malware signature or blocking a connection to a blacklisted IP address. These systems operate on fixed rules: if a specific condition is met, a specific action follows. But alerts that require contextual analysis or judgment still depend on human review.

In practice, this pipeline is under severe strain. The volume of alerts routinely exceeds what security teams can process. One study found that Security Operations Centers (SOCs) receive an average of 4,484 alerts per day, with 67% going unreviewed due to overload.[77] The problem is

---

[76] Nelson et al., "Incident Response Recommendations and Considerations for Cybersecurity Risk Management."

[77] Tariq et al., "Alert Fatigue in Security Operations Centres: Research Challenges and Opportunities."

compounded by false positive rates that in some environments reach as high as 99%, meaning that some analysts spend most of their time investigating events that turn out to be benign[78]

Autonomous cyber operations could make this problem significantly worse. As noted previously (see “Section 2: Detection-in-Depth”), autonomous cyber agents could potentially scale more sophisticated attacks across many targets simultaneously. High volumes of sophisticated or unusual attack attempts, requiring human review, would strain and even overwhelm the existing triage model.

### AI-Automated Alert Analysis and Triage

Frontier AI (such as LLMs and agentic AI) can be used to automate tasks in the alert pipeline that currently depend on human judgment. Where conventional automation executes predefined responses to known threat signatures, frontier AI systems can interpret telemetry in natural language, reason about ambiguous or novel signals, assess context across multiple data sources, and build situational awareness during active investigations.

Commercial vendors are integrating frontier AI into security operations platforms for AI-assisted triage, investigation, and response—both through established platforms, such as CrowdStrike, Microsoft, and SentinelOne, and through AI-native startups such as Prophet Security, Dropzone AI, and Radiant Security. Vendors report meaningful reductions in investigation times, though independent evaluations in operational environments remain scarce. As offensive capabilities scale, however, even AI-augmented workflows may prove insufficient.

Researchers have proposed a further shift toward fully agentic AI systems that can dynamically generate and adapt their own response workflows—moving from reactive task automation to proactive, self-adjusting threat mitigation.[79] This progression may become necessary to maintain defensive capacity against autonomous cyber attacks at scale.

## Detection Value of AI-Automated Alert Analysis and Triage

**AI-automated alert analysis and triage contributes to detection-in-depth by maintaining defender analytical capacity as attack volume and complexity grow.** Without it, the signals generated by other mechanisms risk going unprocessed. There is emerging evidence suggesting two areas where AI-automated analysis could have immediate detection value.[80]

---

[78] Alahmadi et al.,"99% False Positives: A Qualitative Study of SOC Analysts' Perspectives on Security Alarms."

[79] Ismail et al., "Toward Robust Security Orchestration and Automated Response in Security Operations Centers with a Hyper-Automation Approach Using Agentic Artificial Intelligence."

[80] Covino et al. “Driving Defensive Cyber Automation in Critical Infrastructure.” (Forthcoming)

- **Faster triage and prioritization:** Automating the triage and prioritization of alerts could significantly reduce the time between detection and response. In controlled testing, analysts working with AI-assisted analysis completed investigations 45%–61% faster with higher accuracy than analysts working manually.[81]
- **Filtering false positives:** Automating the filtering of false positive reports would likely reduce the volume of alerts a SOC analyst must process. In certain settings, experimental AI agents have reduced alerts requiring human review by roughly half.[82]

### OT/ICS Environments: Additional Detection Challenges

While current deployments of AI-automated analysis tools are concentrated in IT security environments, the need may be even more acute in operational technology (OT) and industrial control system (ICS) settings that underpin critical national infrastructure. OT networks typically have fewer conventional defenses than IT environments, making detection a primary line of defense—yet only 26% of practitioners rate their detection capabilities as highly effective at identifying ICS-relevant threats.[83] Detection in these environments faces distinct constraints: aging systems are difficult to integrate with modern tools, and active intrusion prevention methods common in IT cannot be used because blocking traffic risks disrupting physical processes.[84]

Alert analysis poses an additional challenge. OT alarms monitor both cybersecurity and physical processes simultaneously—the same alert could indicate a cyberattack, an equipment fault, or normal process variation, and distinguishing between them requires combined expertise in both the technical systems and the physical processes they control that few facilities have on staff.[85]

**AI-automated analysis could help operators interpret these ambiguous alerts faster without requiring them to cede operational authority over critical processes**. However, OT applications remain largely experimental, with only 10% of ICS using any form of AI on process data.[86]

## Design and Implementation Considerations

AI-automated alert analysis and triage's value depends on whether these tools can be deployed reliably, securely, and in ways that earn justified trust from the defenders who use them. Several design and implementation challenges must be addressed before these systems are ready for high-stakes environments.

[81] Cloud Security Alliance, "Beyond the Hype: A Benchmark Study of AI Agents in the SOC."
[82] Mohsin et al., "A Unified Framework for Human AI Collaboration in Security Operations Centers with Trusted Autonomy."
[83] Christopher, "State of ICS/OT Security 2025."
[84] Ibid.
[85] Fung et al., "Adopting AI to Protect Industrial Control Systems: Assessing Challenges and Opportunities from the Operators' Perspective."
[86] Ibid.

## High-Assurance

Introducing an AI-driven tool into the detection pipeline can expand the attack surface and introduce new failure modes if the underlying system is not sufficiently aligned to operator intent and reliable, secure, and interpretable.

**Alignment:** AI systems can pursue objectives that diverge from operator intent, particularly when operating with greater autonomy. For example, Anthropic's alignment testing of Claude Mythos Preview found that earlier versions of the model pushed through safety constraints and took excessive measures to complete difficult tasks, occasionally concealing that it had done these things.[87] In a triage context, a system might pursue a measurable proxy, such as reducing alert volume, too aggressively at the expense of the actual goal of accurate detection.

**Reliability:** AI models can produce hallucinated outputs and inconsistent judgments that lead to poor security outcomes.[88] For example, a system that misclassifies malicious activity as benign could give defenders false confidence that threats have been addressed, allowing intrusions to proceed undetected.

**Security:** AI systems have their own unique vulnerabilities that can potentially be exploited by attackers. For example, prompt injections are attacks that exploit an AI model's inability to distinguish between its own instructions and external input, allowing an attacker to embed malicious commands that manipulate AI behavior.[89] These vulnerabilities exist not only at the model level but across the broader system architecture. AI triage systems may also rely on components such as tool integrations and agentic scaffolding that could be exploited as well.

**Interpretability:** For security teams to trust and act on AI-generated outputs, they need to understand why a system reached a particular conclusion. This is not only a technical requirement but a practical one: analysts must be able to validate AI decisions against their own expertise, explain those decisions to management and compliance teams, and defend them during incident reviews. Commercial tools increasingly offer natural language explanations of their decisions, but these are themselves AI-generated and may not faithfully represent the system's actual reasoning. Research on explainability in SOC environments has found that poorly designed explanations can impose cognitive load, create misplaced confidence, or even degrade decision quality.[90] Building triage tools that are genuinely interpretable remains an open challenge.

---

[87] Anthropic, "System Card: Claude Mythos Preview."
[88] Potter et al., "Frontier AI's Impact on the Cybersecurity Landscape."
[89] Vassilev et al., "Adversarial Machine Learning: A Taxonomy and Terminology of Attacks and Mitigations."

Progress on each of these fronts is needed before these tools are deployed widely, particularly in critical infrastructure environments where errors carry physical consequences.

## Real-World Testing and Integration

The effectiveness of AI-automated triage depends as much on how systems are integrated into operational environments as on the capabilities of the underlying models. These tools do not operate in isolation, but need to be embedded in an organization's existing security infrastructure, such as its EDR platforms or incident ticketing workflows.

In many cases, organizations operate fragmented tool environments with incompatible data formats and interfaces, and current AI tools vary considerably in the breadth and depth of integrations they support. In OT/ICS environments, these challenges are compounded by legacy systems, proprietary protocols, and the operational constraints described above. Realizing the detection value of AI-automated triage and analysis will require sustained investment not only in AI capabilities but in the data infrastructure and interoperability standards that allow those capabilities to function across heterogeneous environments.

Also, trust in AI-enabled defensive automation will need to be earned through validation under real-world conditions. Defenders should not delegate decision-making entirely to systems that have not demonstrated reliability in their specific operational context. Early research supports a graduated approach to autonomy. In a simulated cyber-range exercise, analysts working with an AI-assisted SOC agent began by reviewing every output and approving all critical actions; only after the system demonstrated reliability over time did they allow it to handle routine decisions independently, while they focused on more complex threats.[91]

---

[90] Rastogi et al., "Too Much to Trust? Measuring the Security and Cognitive Impacts of Explainability in AI-Driven SOCs."
[91] Mohsin et al., "A Unified Framework for Human AI Collaboration in Security Operations Centers with Trusted Autonomy."

### From Autonomous Threat Analysis to Autonomous Response

While this section focuses on AI-automated alert analysis and triage, a related area of development is autonomous response. These are systems that can independently execute routine response actions such as isolating compromised endpoints, blocking malicious connections, or generating incident tickets without waiting for human approval. Where AI-automated triage helps defenders identify threats faster, autonomous response would allow defensive actions to match the speed and scale of agentic attacks—ensuring that the capacity gains from faster analysis are not lost to bottlenecks in human-executed response.

In the CyberAlly research exercise, a semi-autonomous SOC agent trained on two years of incident data was deployed alongside human analysts. Over the course of the exercise, the system reduced mean time to respond from eight hours to 90 minutes, cut alerts requiring human review by half, and increased automated ticketing from 10%–75% of incidents.[92] These results suggest that autonomous response could eventually complement AI-assisted triage, though the same trust-building and validation requirements apply—and the stakes of autonomous action are higher, since response errors can directly disrupt operations.

## Recommendations

1. **AI companies with advanced cyber-capable models should work with defenders and security researchers to develop and test AI-automated alert analysis and triage tools.**

AI companies possess unique advantages in this space: deep expertise in the frontier models that underpin these tools, the engineering capacity to build and iterate on integrations with diverse security environments, and the ability to provide early or privileged access to state-of-the-art capabilities. But the high-assurance requirements for deploying AI in security operations—alignment to operator intent, reliability under adversarial conditions, robustness to exploitation, and genuine interpretability—cannot be met through model development in isolation. They require sustained collaboration with the defenders and researchers who understand the operational environments, threat landscapes, and failure modes these systems will encounter.

Real-world testing with critical infrastructure operators is essential to determine whether AI-automated triage tools can function reliably under these conditions. In critical infrastructure settings, legacy systems, proprietary protocols, and constraints against active traffic blocking compound integration challenges. This means tools validated in IT environments cannot simply be ported over.

---

[92] Ibid.

Differential access[93] schemes offer a promising vehicle for developing and refining automated alert analysis tools. Anthropic's Project Glasswing, for example, provides vetted organizations with early access to Claude Mythos Preview, a frontier model with advanced cybersecurity capabilities.[94] However, this initiative has focused on vulnerability discovery and patching: using the model to identify and fix zero-day flaws in critical software before adversaries can exploit them. But the same differential access mechanism could be extended to accelerate the development and validation of automated cyberdefense tooling, including alert triage systems tailored to critical infrastructure. Giving OT/ICS security researchers and operators early access to frontier models would allow them to test integration with their specific environments, surface failure modes, and build operational trust before broader deployment.

---

[93] Ee et al., "Asymmetry by Design: Boosting Cyber Defenders with Differential Access to AI."
[94] Anthropic, "Project Glasswing."

# 5 | Agent Honeypots

*Stakeholders: Existing honeypot operators, defenders,  governments, philanthropists. For implementation, see Recommendations.*

**One promising detection mechanism is agent honeypots—decoy systems or resources designed to attract autonomous attackers and reveal their methods.** Honeypots are a well-established tool in cyberdefense. The emergence of AI-driven offensive operations creates both a new need and new opportunities for their use. Traditional honeypots are designed with human attackers or simple bots in mind. Agent honeypots would be specifically designed to detect autonomous offensive cyber operations, using techniques tailored to the distinctive characteristics of agent-driven activity.

Agent honeypots contribute three distinctive capabilities to our detection-in-depth approach. Because a honeypot has no legitimate users, any interaction with it is inherently suspicious, producing high-confidence alerts with near-zero false positives. Because the honeypot is entirely controlled by the defender, they can instrument it to actively probe whether an attacker is an AI agent, using techniques like prompt injection and temporal analysis that exploit how LLMs process information. And because honeypots are involved in real attacks rather than simulated ones, they generate direct empirical data on agent prevalence, tactics, and capabilities.

## What are Honeypots?

### Traditional Honeypots

Honeypots are a well-established tool in cyberdefense and security research. Their use dates back to at least 1986, when Clifford Stoll used a deliberate bait to track a West German hacker infiltrating Lawrence Berkeley National Laboratory.[95] The core idea is straightforward: a defender places a decoy resource on their network, such as a fake server, database, or set of credentials. This resource is supposed to appear valuable to attackers, but is in fact isolated and monitored, with no legitimate users. Any interaction with it is therefore inherently suspicious.

Honeypots take many forms. They are used as research tools to study threats and as defensive mechanisms in production systems. They can sit on the open internet to attract attackers scanning for targets or deeper inside an organization's network to catch attackers who have already gained access. At the simplest end, honeytokens, such as fake credentials or decoy documents, act as lightweight tripwires that alert defenders when accessed. At the other end, high-interaction honeypots run real or near-real operating systems and services, allowing attackers to operate freely within a monitored environment so that defenders can study the full range of their behavior. Large-scale deployments like Amazon's MadPot operate tens of thousands of sensors mimicking cloud workloads, combining threat intelligence gathering with automated defensive response (see Appendix II for detailed examples).

[95] Greene, "Honeypot (computing)."

Honeypots already pick up attacks from automated attacks, such as scanners and botnets. However, such attacks often follow rigid patterns that make them easy to fingerprint. In contrast, AI Agents can vary their tactics, techniques, and procedures (TTPs) to avoid fingerprinting.

## Agent Honeypots

**Agent honeypots would extend traditional honeypot techniques to target autonomous AI-driven cyberattacks specifically.** Such honeypots could extract information that is specific to AI agents, such as system prompts or model identifiers. Agent honeypots could also be designed to be especially attractive for agent-based attackers, such as services with low security offering compute resources that agents need to scale their operations. Further, they can exploit specific vulnerabilities of AI agents to classify and probe attackers. For example, Palisade Research augmented a standard honeypot with prompt injections that were designed to manipulate AI agents into running commands no human or conventional bot would execute (see case study below).[96]

[96] Traditional software bots ignore such injections because they do not process natural language, and human attackers are unlikely to comply with unusual instructions embedded in system output, see Reworr and Volkov, "LLM Agent Honeypot: Monitoring AI Hacking Agents in the Wild."

### Case Study: The Palisade Research LLM Agent Honeypot[97]

Researchers from Palisade Research built a honeypot designed to detect AI agents attempting to break into computer systems. They modified the popular SSH-honeypot Cowrie[98] to include two AI-specific detection techniques. The first, a prompt injection, embedded hidden instructions in the text displayed to anyone connecting to the honeypot. Because AI agents built on large language models read and respond to all text they encounter, these instructions could manipulate them into running a command no human or conventional bot would execute. The second, temporal analysis, exploited the fact that AI agents respond far faster than human attackers, allowing the system to distinguish machine-speed replies from human ones.

As of early 2026, the project had recorded over 21 million connection attempts and identified 14 potential AI agents, of which three were confirmed through both techniques.[99] These low numbers suggest that while autonomous hacking agents exist and are detectable, they are not yet a significant force in real-world cyber operations. However, the researchers noted significant limitations: the honeypot captured only a narrow slice of global attack traffic, most of which was likely opportunistic scanning activity on low-value targets. Additionally, both techniques are new and tested at limited scale, and the system detects only fully autonomous agents, not other forms of AI-assisted attack.

## Detection Value of Agent Honeypots

Agent honeypots could help accomplish a number of goals:

1. **Generating threat intelligence:** Gathering empirical, real-world data about how AI agents are actually being used in cyberoffense, such as how prevalent they are, what tasks they are being directed to perform, what models and frameworks underpin them, and how their operators have configured them.
2. **Detecting ongoing attacks:** Alerting defenders to external attacks and internally suspicious behavior to enable rapid response.
3. **Strengthening defenses and active disruption.** Hardening defenses and patching vulnerabilities based on observed attacks, slowing down attackers, and counterattacking agent operations[100].

---

[97] Reworr and Volkov, "LLM Agent Honeypot: Monitoring AI Hacking Agents in the Wild."
[98] GitHub, "cowrie."
[99] Up-to-date detection statistics are displayed on: AI Honeypot, "LLM Agent Honeypot."
[100] Given the report's focus on *detecting* autonomous offensive operations, we discuss uses related to strengthening defenses and active disruption in Appendix III

These purposes apply across two threat models: *external attacks*, where agents target an organization's systems from the outside,[101] and *insider threats*, where agents deployed within an organization operate outside their intended scope or conduct sabotage.

## Threat Intelligence

The most fundamental contribution of agent honeypots would be empirical, real-world data about AI-driven cyberoffense. At present, most of what we know about offensive AI capabilities comes from pre-deployment evaluations, which measure what agents can do under controlled conditions, but cannot tell us how widely agents are actually being deployed for attacks, what tasks they are performing, which models and frameworks are in use, or how agents behave against real-world targets. Aggregated honeypot data can help policymakers assess the broader threat landscape, AI developers evaluate real-world misuse of their models, and safety researchers study how autonomous systems behave in uncontrolled environments.

**Understanding how agents are used for attacks.** Honeypots allow defenders to observe agent behavior in detail: what commands are run, what tools are deployed on the target, how the environment is explored, and what resources are prioritized. For example, an SSH honeypot could capture login attempts, post-compromise commands scanning for resources and uploaded exploits. This could capture *the degree of human involvement*. For example, timing patterns can help distinguish between agents operating under human direction and those acting fully autonomously (e.g., if it pauses at decision points, it may be more likely that a human operator is reviewing its output and guiding its next steps). Additionally, honeypots could give signals about attacker objectives by revealing which honeypots are targeted and how an agent prioritizes once inside a network. This information also reveals the types of attacks for which agents are most commonly used.

**Identifying technical characteristics of an agent deployment.** Agent honeypots could reveal details about the attacker's technical infrastructure that go beyond what traditional threat analysis can extract. If the agent is using a commercially hosted model that implements text watermarking (such as Google's SynthID-Text[102]), watermark detection on the agent's text output could identify the provider.[103] Prompt extraction attacks could extract an agent's system prompt,[104] revealing the attacker's intent and the operational parameters the agent has been given. In the best case, a honeypot interaction might induce the agent to upload or reveal its own configuration, weights, or scaffolding code, giving defenders complete insight into the goals and technical setup of the system. Finally, defenders could cross-reference attacks on different honeypots to identify larger-scale patterns of coordinated attacks and map command-and-control networks. For example, by cross-analyzing behavioral signatures across multiple incidents.

[101] This includes attackers that have gained network access and are now running exploits and moving laterally.
[102] Dathathri et al., "Scalable watermarking for identifying large language model outputs."

**Assessing attacker capabilities in real-world conditions.** Agent honeypots can be designed with defenses of varying difficulty. The success and speed with which agents overcome these challenges let defenders assess what agents used by real attackers can achieve in practice—something lab-based evaluations cannot capture. Maintained over time, this creates a living measure of how offensive cyber agent capabilities are improving in the real world, which can inform both defensive priorities and policy decisions about AI risk.

**Detecting rogue agent activities.** While more speculative, in the future, autonomous agents could become capable of operating for sustained periods outside of human control, using their capabilities to acquire resources and expand their deployment.[105] This novel type of threat actor could be highly difficult to identify through conventional means. For example, their behavioral patterns would not correspond to established human TTPs. However, they would likely need specific resources to sustain their operation: compute capacity, cloud credentials, cryptocurrency, or access to additional systems. Placing agent honeypots around GPU clusters, cloud credential stores, cryptocurrency wallets, or other resources that a self-sustaining agent would need to acquire could function as one of the few early warning systems against such rogue agents.[106]

**Capturing agent signatures.** Agent honeypots can generate rich interaction logs from which defenders might be able to derive signatures of offensive AI agents. These signatures could be used to attribute different attacks to the same threat actor or feed into defensive systems to flag and block malicious traffic. These signatures range from infrastructure indicators like IP addresses and callback servers to behavioral fingerprints such as exploration behavior, timing of responses, and vulnerability to adversarial inputs. However, agentic signatures may be harder to identify than those of traditional attackers (see Section 1), and their reliability is an open research question.

---

[103] Some recent research on black-box LLM fingerprinting has shown that different models produce statistically distinguishable output patterns, and that techniques such as crafted probe queries can identify specific model versions with high accuracy in controlled settings. Applying these techniques to an agent's output within a honeypot could be possible, though it has not been demonstrated and faces practical challenges: the agent's text may be short, intermixed with tool commands, and transformed by scaffolding code before reaching the honeypot. See Gloaguen et al., "LLM Fingerprinting via Semantically Conditioned Watermarks."
[104] Zhang et al., "Effective Prompt Extraction from Language Models."
[105] Kraprayoon et al., "Highly Autonomous Cyber-capable Agents"
[106] An important, unsolved challenge lies in distinguishing highly autonomous agents that operate under human control and those outside of it.

## Detecting Ongoing Attacks

Honeypots can detect ongoing attacks, thus enabling defenders to rapidly respond to threats posed by AI agents. However, they will only catch the fraction of attacks where attackers happen to interact with a honeypot. Thus, agent honeypots are useful for detection-in-depth, but not a reliable standalone monitoring mechanism.

**Alerts for ongoing attacks.** Agent honeypots can function as early warning systems at multiple layers of an organization's environment. Externally facing honeypots pick up incoming attacks and can alert defenders to operations that appear novel or sophisticated, including autonomous agent activity that might otherwise blend in with high-volume scanning traffic. Agent honeypots deployed inside a network can detect lateral movement by attackers who have already gained initial access, or flag insider threats. Since any interaction with an internal honeypot is suspicious by definition, these alerts are highly informative to defenders and generate near-zero false positives.

**Detecting suspicious insider behavior.** For *insider threat* scenarios, honeypots could also play an important role in detecting suspicious behavior. If an organization deploys AI agents for tasks like IT administration, code review, or data analysis, those agents have defined scopes of operation. A decoy resource placed outside any agent's authorized scope functions as a tripwire: any agent that interacts with it is, by definition, operating beyond its intended boundaries.[107] This could surface cases where an agent has been compromised by an external adversary and redirected toward internal reconnaissance, where a malicious insider has deliberately reconfigured an agent for unauthorized purposes, or where an agent is autonomously drifting beyond its intended scope through misalignment or specification gaming.

## Limitations of Agent Honeypots

Agent honeypots have limitations that shape when and how they can gather relevant information.

**Scope of coverage.** Honeypots require that the attacker interact with a resource the defender controls. This covers most of the attack chain, including reconnaissance against public-facing assets and post-compromise lateral movement, but excludes attacks that bypass defender-controlled resources entirely, such as pure volumetric denial-of-service. For AI-enabled attacks specifically, honeypots are most effective when the agent directly executes operations against the target. When AI is used only in preparatory phases, deception can still be partially effective—fake public information consumed during reconnaissance can propagate detectable artifacts into the resulting attack. But honeypots cannot reach AI systems that operate entirely offline, such as generating generic malware from training data alone.

[107] For example, a misaligned AI deployed internally in an AI company might attempt to exfiltrate its weights to operate outside of the company's control (see Clymer et al., "The Rogue Replication Threat Model."). To catch this, the company could, for example, deploy decoy files that look like model weights, which give alerts when anybody reads, edits, or moves them.

**Detecting and evasion.** If attackers detect that a resource is a honeypot, they will often evade it and redirect their attack elsewhere, in which case little valuable information is gained. Skilled human attackers can often detect and evade honeypots, using techniques ranging from probing how a system handles unusual network traffic to checking whether files and configurations are consistent with a real machine.[108][109][110] Agents trained on cybersecurity data, or instructed by their operators to check for signs of deception, could do the same. This makes successful deception a key design priority, which we discuss in a following sub-section ("Deceive").

**Cost.** Honeypots cost money to develop, deploy, maintain, and analyze the data they produce. Costs can range from a few thousand to millions of dollars (see Appendix IV for estimated costs). Especially custom-built, high-interaction honeypots deployed at large scale can be prohibitively expensive for small actors.

**Spoofing human behavior.** Attackers might attempt to hide the fact that attacks are conducted by an agent by making their behavior indistinguishable from that of humans or scripts.[111] For example, they could slow down their responses to avoid timing-based agent detection. This would make it more difficult to specifically study agent-based cyberattacks.

**Honeypots as attack surfaces.** If honeypots are not properly secured and isolated (e.g., if they share a network segment or credentials with production systems), attackers can use them as a foothold to pivot to real infrastructure. If an agent is aware it is interacting with a honeypot, it could deliberately generate misleading indicators to poison the resulting threat intelligence[112] or flood honeypots with traffic to distract from real attacks.

[108] Srinivasa et al., "Gotta Catch 'em All: A Multistage Framework for Honeypot Fingerprinting."
[109] That said, even skilled professionals can be fooled: highly realistic honeypots (such as perfect clones of production systems) can pass expert scrutiny, and attackers under time pressure, or moving laterally inside a network after initial compromise, tend to be less careful. Some categories of honeypot, such as fake credentials or canary tokens, are inherently harder to detect since an attacker cannot test whether they are real without actually using them.
[110] The largest study on honeypot effectiveness found that 120 out of 120 red teamers exploring a network containing honeypots interacted with honeypots extensively, even if they were warned that honeypots exist in the network. This shows that even professionals still fall for honeypots at high rates. See Ferguson-Walter et al., "Examining the efficacy of decoy-based and psychological cyber deception."
[111] Non-malicious web-using agents are already doing this. Staufer et al. 2026 found that 6 out of 30 studied agents attempted to mimic human web traffic requests by spoofing Chrome-like User-Agent strings and residential IP contexts. Mato, "Browser agent bot detection is about to change" explicitly markets that their agents can bypass anti-bot systems.

## Design and Implementation Considerations

Designing an effective agent honeypot requires navigating design decisions, which we organize into four categories:

1. **Attract:** How to get the relevant agents to encounter and engage with the honeypot, through choices about where it is deployed, what it emulates, and how it is made discoverable.
2. **Deceive:** How to prevent the agent from recognizing it is interacting with a honeypot, so that it continues to engage and behaves as it would against a real target.
3. **Gather:** How to collect useful data from the interaction and draw inferences about the agent's behavior, capabilities, technical setup, and objectives.
4. **Integrate:** How to use the gathered information, for example, by feeding it into security operations, correlating it with other intelligence sources, and sharing it across organizations.

The right choices for each category depend on the threat model (e.g., external attack versus insider threats), the strategic purpose (e.g., threat intelligence, early warning, or enabling disruption), and the affordances of the actor implementing the honeypot (e.g., technical capabilities or which services they can credibly emulate). The design space for honeypots is large, and honeypots optimized for different uses can look quite distinct from one another (see Appendix VI for differences in agent honeypot designs).

### Attract

An agent honeypot is only useful if the relevant agents encounter and engage with it. Agent honeypots must therefore be discoverable by agents and attractive enough to warrant attack.

The most fundamental choice is placement. For externally facing honeypots intended to capture offensive agents operating on the open internet, the system must be discoverable through the channels that these agents and their operators use to find targets. In practice, this might mean being indexed by search engines or appearing as the kind of exposed service that vulnerability scanners flag. For internal honeypots, the decoy must sit where an agent operating outside its scope or an attacker moving laterally would plausibly encounter it, but where legitimate users would have no reason to access it.

[112] Palo Alto Network describes a case where attackers fed domains associated with legitimate security companies into research honeypots, leading those domains to be blocked by other security operators (See: Palo Alto Networks, "What Is a Honeypot?").

The type of system a honeypot emulates also shapes what kinds of agents it will attract. Emulating a commonly targeted internet-facing service like SSH or a web application will capture broad, opportunistic scanning activity, including whatever proportion of that traffic is driven by AI agents. To attract a larger share of sophisticated AI agents, one can emulate high-value targets, such as industrial control systems or military communication networks. To attract attacks from rogue agents, honeypots could emulate compute resources, model weights, or services such as banking protected by know your customer/client (KYC).

## Deceive

Once an agent has encountered the honeypot, the next challenge is to make it believe that the honeypot is real. If the agent recognizes it is interacting with a decoy, it will disengage, alter its behavior, and alert its operator, resulting in the honeypot losing its value. The design goal is a resource convincing enough that the agent behaves as it would against a real target.

Initially, agent honeypots must balance attractiveness and realism. If a honeypot looks too good to be true, this could raise suspicion, and agents might avoid engaging with it. A server with default credentials and known unpatched vulnerabilities will attract attention, but sophisticated attackers will recognize that real targets have better security. Further, the honeypot must respond just like real systems while the agent is performing reconnaissance. Otherwise, attackers can send packets to check for known honeypot behaviors[113][114] or analyze traffic of the honeypot[115] to determine if a resource is a decoy.

During the attack, honeypots need to maintain the deception. If the agent notices unrealistic responses, suspicious traffic, logging infrastructure, or known honeypot signatures, it will disengage. Honeypots become harder to detect with greater interaction depth, since attackers must explore more before reaching the boundaries of the simulation. Custom-built environments are also more resistant, since attackers cannot match them against known signatures of widely deployed platforms. The least detectable honeypots closely mirror real systems: they run on dedicated hardware instead of a virtualized environment, clone actual production environments and traffic patterns, and conceal their logging infrastructure. Honeytokens, such as fake credentials, are particularly resistant to detection, since an attacker cannot test them without triggering the alert.

Deception might become harder as agent frameworks mature and AI cyber capabilities improve. Agent scaffolding may include automated checks for known honeypot indicators, such as a tool call that checks Shodan's honeyscore[116] of the targeted IP address. AI agents might gain new capabilities that enable them to better circumvent honeypots. And operators may instruct their agents via their system prompt to test the target system and look for signs that it is a decoy. On the other hand, defenders could respond by using AI to improve their deception techniques[117]. First, LLMs can be integrated directly into honeypots to generate dynamic responses to attacker commands. For example, an SSH honeypot where responses are LLM-generated can provide deeper interaction than static outputs.[118] Second, AI tools such as coding assistants may accelerate the development and deployment of honeypots themselves, thus reducing the time and expertise required to stand up realistic deception environments.[119] Thirdly, defenders could use their own AI agents to test whether they can distinguish their honeypots from real resources. If these approaches mature, defenders could rapidly spin up and test customized and realistic honeypot environments tailored to specific threats or vulnerabilities as they emerge, which may help them keep better pace with developments on the offensive end.

---

[113] Vetterl and Clayton, "Bitter harvest: Systematically fingerprinting low-and medium-interaction honeypots at internet scale."
[114] Srinivasa et al., "Gotta Catch 'em All: A Multistage Framework for Honeypot Fingerprinting."
[115] If a system has no in- and outgoing traffic or if the traffic follows a simple pattern, it is likely to be a decoy.

## Gather

Agent honeypots should be designed to gather useful information about attacking agents. Thus, their design depends on the strategic objective for the agent honeypot. A honeypot focused on detection needs fast, reliable signals that an attack is underway. A honeypot focused on threat intelligence needs to observe how an attack unfolds from start to finish, which may mean letting an interaction run longer before intervening. A honeypot intended to support disruption needs enough information about the agent's architecture to attempt adversarial techniques against it.

These objectives create tradeoffs in how aggressively data is collected and how long an agent's interaction with a honeypot needs to be. More aggressive data collection, such as prompt injections, can reveal more information, but also risk the agent disengaging from the interaction. Additionally, depth of interaction trades off cost against the amount of collected data. A shallow, low-interaction environment that emulates only surface-level services (such as a login prompt) is cheap to deploy, but only captures limited data. A deep, high-interaction environment that runs real or near-real services captures far richer data but is significantly more expensive to build and maintain.[120]

---

[116] Shodan is a search engine for internet-connected devices that provides a "honeyscore" API endpoint that rates the probability (0–1) that a given IP address is a honeypot based on service fingerprints and hosting metadata: GitHub, "metasploit-framework."

[117] This dynamic already exists between honeypot operators and sophisticated human attackers, but it may accelerate in the agent context: once a honeypot detection technique is developed and incorporated into a widely used agent framework, it is immediately available to every operator using that framework, unlike human tradecraft which diffuses more slowly.

[118] Some recent projects have demonstrated this approach. Splunk's DECEIVE proof-of-concept uses an LLM backend to simulate an entire Linux server via SSH, generating realistic command outputs without running a real operating system; the system can be reconfigured to emulate different types of targets simply by updating the AI prompt, see Bianco, "Introducing DECEIVE: A Proof-of-Concept Honeypot Powered by AI."

[119] Beelzebub, an open-source honeypot framework, uses LLMs to create high-interaction honeypot environments that deploy in minutes via a single configuration file, supporting multiple protocols from a single instance, see Beelzebub, "Open-source AI-powered honeypot framework."

The raw observational data collected by honeypots include network metadata, authentication attempts, the full sequence of commands run, interactions with available services and any uploaded artifacts. From this, defenders can reconstruct how an attack unfolded and what its apparent goal was, whether stealing data, establishing persistent access, or acquiring compute resources.

A key priority across all uses of agent honeypots is to distinguish AI agents from human attackers or bots, so that resulting intelligence is specific to agentic threats. Several approaches are possible, but most lack experimental validation against advanced agents that attempt to evade detection. Honeypots can exploit differences in how LLMs process text: encoding information using special characters that LLMs read differently from humans,[121] embedding prompt injections that trigger specific responses in AI agents,[122] or presenting large volumes of text that only a machine could process quickly. Response timing[123] and broader behavioral patterns—such as how the attacker explores the environment or adapts after failed commands—provide additional signals. Reliable classification will likely require combining multiple such indicators.

Finally, how broadly honeypots are deployed shapes what conclusions can be drawn. By aggregating data from many honeypots, it is possible to uncover coordinated campaigns across organizations, shifts in agent capabilities over time, or infrastructure being reused in many attacks. This is a resource allocation decision: broader coverage yields richer intelligence but requires more infrastructure and analytical capacity.

## Integrate

The intelligence gathered by agent honeypots is only as valuable as the defender's ability to act on it. To achieve this, gathered data should be integrated with defensive systems and processes. Alerts from honeypots should feed into the organization's existing security operations workflow so that network defenders can respond rapidly and appropriately. They might also trigger automatic responses, such as blocking certain IP addresses from using services via a firewall or restricting permissions of internal agents via identity and access management systems.

---

[120] Costs of honeypots range from a few thousand dollars for a pre-built, low-interaction agent honeypot to tens of millions for enterprise-scale deployments (see Appendix IV for estimated costs).
[121] Ayzenshteyn et al. (2025) use special unicode characters.
[122] Volkov and Dmitrii (2024) as well as Heckel and Weller, "Countering autonomous cyber threats" use prompt injections to distinguish AI agents from humans and regular bots.
[123] Timing-based analysis is used to distinguish human and agent-based attackers in: Reworr and Volkov, "LLM agent honeypot: monitoring AI hacking agents in the wild."

Individual honeypot deployments capture narrow slices of the overall threat landscape. Pooling data across organizations enables the identification of trends, emerging capabilities, and coordinated campaigns that no single defender could see. Potential institutional mechanisms needed to support such sharing are discussed in Section 6: "An Agentic Cybersecurity Exchange (ACE)". Finally, insights from honeypot operations should feed back into honeypot design itself: if certain attack patterns recur, honeypots can be reconfigured to capture more detail. If agents are found to be detecting certain deception techniques, those techniques need updating.

## Who Could Develop and Deploy Agent Honeypots?

Agent honeypots could be developed and deployed by a range of actors, each with different capabilities, incentives, and threat models (see Appendix VII for a detailed discussion):

- **Frontier AI companies** are uniquely positioned to deploy honeypots against threats from their own models, such as a misaligned model attempting to exfiltrate its own weights or escalate its privileges.
- **Cloud compute providers** like AWS, which already operates the MadPot honeypot network, have the infrastructure to deploy at massive scale, and the commercial incentive to understand threats facing their networks. Additionally, they can develop highly realistic honeypots that emulate compute environments to attract attacks from autonomous agents.
- **Critical infrastructure operators** control the network segments where industrial control system honeypots need to be placed, though they often lack the in-house capability to build them.
- **Security researchers and threat intelligence companies** have the expertise to develop novel honeypot designs and the motivation to identify emerging threats early, as the Palisade Research project demonstrates.
- **Intelligence and national security agencies** have the resources, mandate, and analytical capacity to deploy large-scale honeypot networks, and could fund programs that individual organizations cannot sustain alone.

In many cases, the most effective approach will involve partnerships between those who build novel honeypot designs and those best positioned to deploy and maintain them at scale.

## Recommendations

1. **Governments and philanthropists should fund public research and development into agent honeypot design and detection techniques.**

The most pressing gap is empirical. We do not yet know which agent detection techniques will be reliable against increasingly sophisticated systems, which honeypot designs generate the most operationally useful intelligence, or whether the approaches that work today will continue to work as agent frameworks mature. Closing this gap requires R&D investment in both building honeypots and systematically testing them against a range of agent architectures and capability levels. Governments, through agencies like the Cybersecurity and Infrastructure Security Agency (CISA), Defense Advanced Research Projects Agency (DARPA), or their international equivalents, are well-positioned to fund this work. The results are a public good that benefits the entire security ecosystem, and the commercial incentive for any single firm to invest is limited at this early stage of the field. Priority areas include developing and evaluating a diverse repertoire of agent classification techniques, testing deception techniques against agents of varying sophistication, and building higher-fidelity honeypot environments that can sustain extended interactions with capable agents.

2. **Existing honeypot operators should begin collecting data on agent activity.**

There is very little real-world data on AI agents conducting cyberattacks, and gathering this data should be an urgent priority. Organizations that already operate honeypots should begin augmenting them with even basic agent detection capabilities to start distinguishing agent activity from conventional attacks. The resulting data does not need to be perfect to be useful: even rough estimates of the prevalence and characteristics of agent-driven attacks would represent a significant advance over the current evidence base, which consists of a handful of research deployments. Threat intelligence companies, cloud providers, and security researchers are the most natural early movers, since they have existing honeypot infrastructure and the analytical capacity to make sense of the data.

3. **Governments and philanthropists should create market incentives for private-sector development of agent honeypot products.**

A commercial deception technology market with $2.7B annual revenue already exists.[124] Adapting these products to detect autonomous agents is likely the fastest path to broad deployment of agent honeypots. But most potential customers do not yet perceive agent-driven attacks as a near-term threat, so vendors lack the market signal to invest. Governments can pull development forward by acting as early customers of agent honeypot products and the threat intelligence produced by them. Grant programs could also support early-stage companies and open-source projects building agent detection into existing honeypot platforms.

---

[124] The market size is estimated at $2.7B in 2026, of which a significant part is likely honeypots. See Mordor Intelligence, "Deception Technology Market Size & Share Analysis - Growth Trends and Forecast (2026 - 2031)."

# 6 | Agent Security Alert Standard

*Stakeholders: AI Industry consortia, trusted national or independent standards bodies. For implementation, see Recommendations.*

Security alerts, variously known as advisories, bulletins, and vulnerability notes, are technical notifications that inform the ecosystem about active threats, enabling defenders to detect, disrupt, and prepare against them.[125] The field of agentic cybersecurity has only recently begun producing such alerts, with early examples including OpenAI's "Disrupting Malicious Uses of AI" report series[126] and Anthropic's threat alert for the GTG-1002 campaign.[127]

Such public alerts complement centralized sharing and coordination systems, including our recommended Agentic Cybersecurity Exchange (see Section 7: Agentic Cybersecurity Exchange (ACE) ), which face significant coverage gaps in practice. Often, participation in centralized threat-sharing infrastructure is deterred due to the meaningful overhead and expertise required.[128] Even among organizations that do participate, shared data can be difficult to transform into threat detection, as it is often high-volume, under-analyzed, and lacking in context.[129] For small- and medium-sized organizations, the result is a systematic gap in threat awareness.[130] A further challenge is global reach. Often, these institutions are scoped domestically or among small groups of like-minded nations, limiting the impact of intelligence.[131] Public alerts can bypass these barriers, delivering pre-analyzed, actionable intelligence openly and globally to any organization that needs it, regardless of capacity.

Existing agent threat alerts, however, fall short of this potential. First, threat signature data is reported inconsistently: some reports enumerate TTPs systematically,[132] others embed them loosely in threat narratives,[133] and specific threat indicators are rarely included, limiting actionability. Second, model and agentic methods disclosure is thin; reports often cite product names without specifying model versions[134] and exclude methodologies, including orchestration frameworks and jailbreaking techniques. This limits both the value of alerts for signature development and building general situational awareness of threat trends. Third, narrative claims are often definitionally fuzzy or unsubstantiated: attribution assertions often lack evidentiary explanation, while quantitative claims about the level of AI involvement are difficult to evaluate without agreed definitions.[135] Fourth, reporting is often selective and slow: OpenAI publishes every few months, Anthropic's GTG-1002 report appeared long after the campaign was active,[136] and selective pre-publication sharing with unspecified "partners,"[137] often at organizational discretion, can mean many ecosystem actors receive no timely warning at all.[138]

---

[125] Johnson et al., "Guide to Cyber Threat Information Sharing."
[126] Nimmo et al., "Disrupting malicious uses of AI: October 2025."
[127] Anthropic, "Disrupting the first reported AI-orchestrated cyber espionage campaign."
[128] Abraham et al., "Promoting research on cyber threat intelligence sharing in ecosystems."
[129] Dykstra, "Maximizing the benefits from sharing cyber threat intelligence by government agencies and departments."
[130] Wilburn et al., "Cyberattacks are hurting US businesses. Here's how Congress can upgrade cybersecurity information sharing."
[131] Our ACE recommended in Section 7 is intentionally designed around these challenges. It aims to be global and oriented towards open reporting and public security. Still, the ACE will have its limits and if it is not implemented, today's cybersecurity institutions are ill-suited for wide distribution of threat alerts. Agent security alerts, and this standard, should be implemented to circumvent these limits.
[132] Nimmo et al., "Disrupting malicious uses of AI: an update."
[133] Google Threat Intelligence Group, "GTIG AI Threat Tracker: Distillation, Experimentation, and (Continued) Integration of AI for Adversarial Use."

To resolve these reporting challenges, we recommend that agentic infrastructure providers and/or standards bodies establish a unified agentic security alert standard. As agent security alerts are still nascent, and reporting norms established now through practice will be difficult to change later, it is important to act promptly before ad hoc habits calcify into de facto standards. The standard we are proposing should serve the following goals:

- **Ecosystem-wide broadcast:** It must ensure timely, public communication of detected threats so defenders across the ecosystem can respond.
- **Disruption and defense enablement:** Alert content should include actionable intelligence and signature data to enable disruption and defense.
- **Capability assessment:** Alerts should seek to provide for greater situational awareness of threat actors, methods, and capabilities to improve defensive strategy and help policymakers benchmark this emerging threat.

The primary actors issuing these alerts will be agent infrastructure and model providers, given their unique window into model misuse and understanding of the threat environment.

---

[134] Recent reports from Anthropic and Google simply city “Gemini” and “Claude Code.” See Anthropic, “Disrupting the first reported AI-orchestrated cyber espionage campaign” and Google Threat Intelligence Group, “GTIG AI Threat Tracker: Distillation, Experimentation, and (Continued) Integration of AI for Adversarial Use.”

[135] For instance, Anthropic’s threat alert about GTG-1002 did not substantiate their attribution of Chinese state-sponsored actors. Further, they did not define how their claim of 80–90% AI automation was measured or if those figures included the potentially substantial human effort required to develop the attack methods. See Anthropic, “Disrupting the first reported AI-orchestrated cyber espionage campaign.”

[136] Anthropic, “Disrupting the first reported AI-orchestrated cyber espionage campaign: Full Report.”

[137] Nimmo et al., “Disrupting malicious uses of AI: October 2025.”

[138] The costs of this failure are compounded by two features of the agentic threat environment. First, models may be fungible within attack frameworks, meaning disrupting a threat at one platform does not eliminate it if other providers remain unaware and unprotected. Second, agentic attack methods may evolve rapidly once publicly disclosed, creating a narrow window in which a report is maximally actionable before adversaries adapt. Both dynamics make timeliness not merely a best practice but a core functional imperative of any effective alert standard.

# Design and Implementation Considerations

A complete agentic security alert standard requires the following components:

## Governance Standards

This component should seek to establish both the language and practice of the standard. This should include:

- **Common definitions** for key terms such as "AI agent" and "degree of attack automation" to ensure clarity and ecosystem trust;
- **Availability and Broadcast expectations**, including human- and machine-readable formats,[139] and expectations regarding proactive distribution to affected parties and those capable of attack disruption;
- **Timeliness expectations** informed by coordinated vulnerability disclosure. These expectations must be further adapted to agentic threat dynamics, such as the possibility that agent threats may also read threat alerts and adapt tactics.

## Alert Content Standards

Reports should meet the general threat bulletin essentials defined by NIST,[140] listing affected platforms, estimated impact, severity rating, mitigation options, references for more information, and alert metadata. Agentic security alerts should supplement these with content tailored to the specific demands of agentic detection:

**Model Use and Operational Architecture:** Reports should detail the specific model(s) and model checkpoints used in an attack to enable model-specific signature development. They should also document operationally relevant architecture details, including elicitation and jailbreaking techniques, orchestration frameworks, any noted downstream agentic hijacking, and the use of agentic swarms where applicable. Finally, reporting should detail the degree of agentic automation.

**Threat Intelligence Indicators:** The indicators included in agentic security alerts should be selected with the understanding that cyber-offensive agents will be uniquely capable of shapeshifting their methods. The "Pyramid of Pain"[141] offers a useful organizing framework: it ranks indicator types by how costly and difficult they are for an adversary to change, with IP addresses and file hashes at the easily-changed base and tactics, techniques, and procedures at the costly-to-change apex.[142] Because capable agents will be able to modify low-level indicators

[139] Machine readable formatting will be essential as agent threats will operate at machine speeds.
[140] Johnson et al., "Guide to Cyber Threat Information Sharing."
[141] Adams et al., "Summiting the Pyramid: Bring the Pain with Robust and Accurate Detection."
[142] Center for Threat Informed Defense, "Detection Engineering Work Overview."

rapidly and intelligently, a well-designed alert standard should weight its indicator requirements toward the pyramid's apex, prioritizing the indicators that are most resistant to agentic adaptation. With this principle in mind, the standard should address the following indicator classes:

**Indicators of Compromise (IOCs):** IOCs, including file hashes, IP addresses, domain names, and similar artifacts, should be assigned low priority. Their detection value has diminished substantially as attacks increasingly avoid leaving static signatures: in 2025, 79% of detected attacks were malware-free, relying instead on living-off-the-land techniques that produce few or no traditional IOCs.[143] Capable agents will only accelerate this trend. That said, IOCs retain marginal value for enriching ecosystem intelligence, contextualizing TTPs,[144] and improving remediation speed,[145] and should remain within scope of the standard on that basis.

**Indicators of Behavior (IOBs):** IOBs are an emerging indicator class that seeks to capture discrete behavioral signatures. These include attack trajectories, chains of activity, and high correlations between otherwise disparate signals.[146] Unlike IOCs, behavioral indicators may be considerably more resistant to modification, since changing behavior requires changing the underlying logic of an attack rather than simply rotating an IP address or recompiling a binary. The field of IOB definition and classification is still developing[147] and will require further investment before IOBs can be reliably shared and operationalized. Still, their potential as agentic signatures makes their inclusion in a standard important even at this early stage.

**Tactics, Techniques, and Procedures (TTPs):** TTPs sit at the apex of the Pyramid of Pain and should be the primary focus of agent security alert standards. They provide a well-developed, common language for describing adversary behavior across organizations, and are the most costly and difficult indicator class for an adversary to change.[148] Intriguingly, certain TTPs produce durable "invariant behaviors," actions that are structurally necessary to achieve an attack objective and therefore cannot be changed without abandoning the objective itself.[149] If an agent must use invariant TTPs', that may act as a durable signature.

**Tool Use:** The tools an agent employs may prove to be a durable and practically useful identifying signature and should also be a high priority focus of an agent security alert standard. Research suggests tool-specific traffic patterns, and patterns of tool invocation, can act as an identifying fingerprint of agentic activity.[150] As tools will be harder for agents to vary,[151] tool-use indicators may be relatively durable signatures to aid detection across the ecosystem. Where tools are centrally managed or logged, tool reporting can also directly support disruption efforts.

---

[143] Crowdstrike, "CrowdStrike 2026 Global Threat Report."
[144] Arctic Wolf, "Understanding Indicators of Compromise and Their Role in Cybersecurity."
[145] Baker, "Indicators of Compromise (IOC) Security Explained."
[146] Frick, "Introducing the Indicators of Behavior (IOB) Sub-Project" and Freed, "Indicators of Behavior and the Diminishing Value of IOCs."
[147] Open Cybersecurity Alliance, "Indicators of Behavior."
[148] Adams et al., "Summiting the Pyramid: Bring the Pain with Robust and Accurate Detection."
[149] Center for Threat Informed Defense, "Level 5: Core to Sub-Technique or Technique."

### Limitations and Uncertainties

As offensive cyber agents are just emerging, basic concepts are often ill-defined,[152] as are the indicators and data needed for appropriate defenses and signature development. While we recommended prompt action on standard setting, actors must be prepared to amend and alter that standard as our understanding grows. There is also uncertainty about whether the industry will share standardized information. Industry has sometimes resisted institutionalized cyber information-sharing efforts, and it is unclear whether any such standard would be embraced, only unevenly adopted, or ignored.

## Recommendations

1. **AI industry consortia should promptly lead agent security alert standards development.**

This standard must both be developed and adopted promptly, while its steward must have a flexible hand to adapt the standard as understanding of agent signatures evolves. To provide nimble interim leadership, private consortia representing key agentic infrastructure and model providers should lead development and commit to frequent standard updates as threats evolve. Given its large organizational membership and stewardship of standards, such as the Model Context Protocol and Agents.MD, the Agentic AI Foundation may be best positioned to ensure wide adoption. The Frontier Model Forum (FMF) is a promising alternative given its existing focus on AI security and information sharing. The FMF's reach may be limited, however, as its membership is both small and exclusively American.

These AI Industry consortia, however, are both limited in reach and have yet to establish global trust. In the medium run, high-reputation standards bodies, such as the National Institute of Standards and Technology (NIST), the Forum of Incident Response Teams (FIRST), or other trusted independent and national standards bodies, may consider further developing and formalizing these standards to ensure global adoption.

---

[150] Zhang et al., "Exposing LLM User Privacy via Traffic Fingerprint Analysis: A Study of Privacy Risks in LLM Agent Interactions."

[151] The tools that are both available and suited to a given attack purpose may be limited. Further, certain tools may demand significant computational resources or inference-time learning to deploy effectively. Finally, within compromised network environments, available toolsets will be deeply constrained.

[152] As Anthropic recently noted in a position paper "The term "AI agent" does not yet have a rigorous, settled definition." For a standard, such definitions are necessary, but so too is the understanding that any definitions must evolve rapidly. See Anthropic, "Request for Information: Security Considerations for Artificial Intelligence Agents Docket No. NIST-2025-0035."

# 7 | Agentic Cybersecurity Exchange (ACE)

*Stakeholders: Governments, agent service providers (AI companies, cloud providers, other agent ecosystem service providers), philanthropists and research funders. For implementation, see Recommendations.*

Cyber-capable AI agents will be able to target multiple entities simultaneously, distribute activity across vendor ecosystems, and leave fragmented signatures that evade traditional detection systems. As these capabilities mature, cybersecurity will increasingly depend on detecting offensive cyber agents operating across the broader digital ecosystem rather than within any single organization or sector.

To address this challenge, we propose an **Agentic Cybersecurity Exchange (ACE), a threat-specific coordination institution dedicated to detecting and disrupting cyber-offensive AI agents.** Led by industry, the ACE would coordinate key agentic infrastructure and service providers, including AI developers and cloud platforms, whose systems can both enable and observe agent activity. By aggregating signals from these actors, the ACE could detect malicious agent campaigns that would otherwise remain invisible to any single organization. Unlike sectoral Information Sharing and Analysis Centers (ISACs), the ACE would operate with a public-security mandate—protecting not just its members but the broader digital ecosystem. Further, the ACE will have a global focus matching the borderless nature of the threat.

The ACE concept draws inspiration from the Global Signal Exchange (GSE), a transnational signal fusion center and disruption coordination body for fraud and scams launched by Google in 2024.[153] The GSE demonstrates that an industry-led, threat-specific, and global-focused organization like an ACE can be viable and effective at scale.[154] More specifically, the GSE demonstrates that critical AI agent service providers, including Google, Meta, Amazon, and Microsoft, are indeed willing and able to build new global institutions to tackle critical problems.[155]

## Core Functions

In service of the coordination function of detection-in-depth approach, the ACE will operate three core functions:

---

[153] Storey and Zebarjadi, "The new Global Signal Exchange will help fight scams and fraud."
[154] Since launch, GSE have grown from analyzing and sharing 40 million signals to over 700 million signals, with approximately 60% of shared signals proving net-new to recipient organizations.
[155] Salmon, "Craig Newmark Backs Global Signal Exchange Anti-Fraud Push."

- **Collect:** The ACE would serve as the digital ecosystem's primary clearinghouse for agent threat intelligence. It would aggregate, enrich, and analyze signals from member organizations and external partners to identify patterns of malicious agent activity that would remain invisible to any single actor.
- **Communicate:** The ACE would provide a central coordination point for receiving agentic threat reports from partner institutions and governments. It would also issue alerts and route threat indicators to affected actors and sectoral information sharing bodies.
- **Act:** The ACE would coordinate disruption of and defense against malicious agent activity across member organizations. While outside the scope of this detection-focused report, the institution should be designed from the outset to enable these coordinated action capabilities.

## Defining Characteristics

The ACE's design is guided by three defining characteristics: a narrow focus on agentic cybersecurity, a public-security mandate, and a global focus.

### An Agentic Cybersecurity Focus

The first defining feature of the ACE is its **narrow focus on agentic cybersecurity** and, more specifically, detecting and responding to offensive cyber agents.[156] As described (Section 1: Introduction), emerging evidence suggests that **offensive cyber agents will be a keystone cyber threat**, given substantial advantages in attack speed, scale, cost, and strategic autonomy. In the near future, it is probable *cyber threats will become* nearly synonymous with *offensive cyber agents*.

An institution tightly focused on this keystone threat is therefore warranted to mobilize the actors, technical expertise, analytical tooling, and operational attention required to respond. Through threat specificity, the ACE will be better able to sustain focus on a problem whose pace and complexity may exceed the bandwidth of other security organizations while concentrating resources toward the specialized tools[157] and expertise needed for detection and mitigation.

Critically, **this focus also allows the ACE membership to be organized around the topology of agentic cybersecurity threats**. Agentic cyberattacks will increasingly operate across model providers, cloud platforms, and software ecosystems spanning multiple industries. Effective detection and disruption will therefore require coordination among a bespoke collection of significant agent service and infrastructure providers.

---

[156] The ACE envisioned in this report is intentionally designed with a narrow mission: to detect and disrupt malicious agents. This constraint is purposeful, because as a mandate expands, focus and resources are inevitably stretched thinner. If resources permit, however, the body could grow over time into a centralized AI cybersecurity hub, assuming further *defensive* functions such as researching the security *of* agents and housing coordinated defensive agent activities, e.g., Project Glasswing, other vulnerability-discovery efforts.
[157] To illustrate why specificity matters, it is likely that agent threat intelligence will contain highly sensitive data such as chat logs. Data management, may therefore demand task specific operational needs such as privacy preserving technologies and bespoke data minimization policies.

Traditional coordination bodies illustrate the limitations of institutions designed around other organizing principles. ISACs, for instance, organize around *industrial sectors*, a structure poorly suited to this threat that operates *across* sectors.[158] ISACs further distribute attention and resources across diverse collections of threats and initiatives, limiting their ability to sustain focus on rapidly evolving risks. The *ISAC-like* Frontier Model Forum illustrates these constraints.[159] Its membership definition, limited to organizations "with the expertise and resources to develop or deploy at scale frontier AI models,"[160] structurally excludes cloud platforms and other infrastructure providers whose telemetry may be essential for detecting malicious agents. Its workstreams, meanwhile, span cyber-enabled AI security, biosecurity, and frontier AI policy, potentially ensuring that no single threat category commands the sustained resources that the pace of offensive cyber agents will likely require.[161]

## A Public Security Mandate

The ACE's second defining principle is that it operates in the interest of public security. This mandate is essential as responsibility for detecting and responding to malicious agents is unevenly distributed: different actors are positioned to observe different dimensions of agentic threats. Agentic infrastructure providers, including model developers and cloud platforms, have unique visibility into model usage, compute workloads, and agent orchestration activity. **These vantage points enable what might be called *origin-point detection*: identifying malicious agent behavior at its source, before attacks propagate to downstream targets, rather than only at the victim's perimeter, where today's detection stack predominantly operates.** These actors therefore have a unique responsibility to the public to detect threats, share threat intelligence, and disrupt attacks when possible.

Once again returning to the ISAC comparison, this public orientation inverts the logic of the traditional ISAC model, which organizes members primarily to defend *themselves* against shared risks.[162] By contrast, the ACE organizes its members to defend the *broader ecosystem,* mobilizing

---

[158] National Computer Emergency Response Teams (CERTs) likewise organize themselves in ways unsuited to the task at hand. Specifically, they organize around national borders, protecting nations from cyber threats, a structure poorly suited for the global nature of these threats.
[159] The FMF is technically not an ISAC, but is *ISAC-like,* given its focus on sectoral security and broad mandate to cover a range of potential hazards.
[160] Frontier Model Forum, "Annual Report."
[161] Frontier Model Forum, "Annual Report."
[162] We address the ISAC model specifically in this section because an "AI-ISAC" has been proposed as a possible solution to these challenges, including in the United States AI Action Plan. The plan's stated goal to "promote the sharing of AI-security threat information and intelligence across U.S. critical infrastructure

those with the visibility and operational capacity to detect and respond to threats that could harm the public.

### A Global Focus

The final defining principle of the ACE is its global focus. This focus is necessary to match the borderless nature of the threat and agent detection. AI is a global enterprise; malicious agents will interact with systems and agents of all nations, and the data necessary to identify these threats will be globally distributed. Only if the ACE takes a global view can it be truly effective. This scope mirrors the Global Signal Exchange whose success is rooted in its ability to collect signals globally.

## Design and Implementation Considerations

When considering the design of an ACE, decision-makers must consider the following:

### Governance and Public Mandate

If the ACE is to serve a public security function, that mandate may need to be structurally secured. The most reliable mechanism would be a public-private partnership combining an industry-operated ACE with fiscal sponsorship from one or more national governments. Such sponsorship would both resource the institution and anchor its public-interest orientation through contracting. The most natural sponsor is the United States. The U.S. hosts many of the world's largest cloud infrastructure providers and frontier AI developers, giving it a uniquely central position in the emerging agent ecosystem. Further, the U.S. already administers foundational elements of global cybersecurity infrastructure, such as the National Vulnerability Database. In view of ongoing developments in the United States regarding cybersecurity program funding, however, policymakers should also consider other models, including sponsorship by a trusted middle power or a multilateral funding structure. Determining the most durable governance model for an ACE remains an open design question and warrants further dedicated analysis beyond this report's scope.

An alternative approach is the membership-driven, non-profit model used by the Global Signal Exchange. To sustain operations, GSE services require a paid membership with only select actors, such as National Police Forces, receiving free services.[163] The benefit of this proven, fee-based model is sustainability and scalability. Still, any such fee-gated information access risks limiting the public security potential of the ACE.

### Institutional Structure

---

sectors," however, is better suited to the ACE model rather than ISAC model's focus on all-hazards and sectoral security mandate. See "Appendix VIII: A Note on the Proposed United States AI-ISAC." See The White House, "Winning the Race: America's AI Action Plan."

[163] Global Signal Exchange, "Who Is Eligible to Join, and More about Free and Sponsored Services."

A core design question concerns the degree of institutional centralization. Three options are worth consideration:

- **A Centralized ACE:** This model offers significant advantages: centralization ensures broad ecosystem coverage, and economies of scale reduce operational overhead. Centralization also fosters a sense of authority and simplifies threat communication.
- **Federated ACE Network:** A system in which a central analytical authority coordinates multiple regional or sectoral nodes that share telemetry and analysis. This structure could balance centralized analytical capability with distributed governance.
- **Regional or National ACEs:** A decentralized model of independent national or regional ACEs, ideally coordinating through shared standards or protocols. This option is modeled on existing CERT networks like CERT-EU.[164]

Each model carries trade-offs. Highly centralized systems improve analytical quality and response coordination but may face legal and political constraints, and risk being single points of failure. More decentralized systems may be politically feasible but risk fragmentation in data quality, inconsistent threat analysis, and slower decision-making.

## Membership

The ACE's detection and mitigation capabilities hinge on whether its members cover the agentic infrastructure stack. While some attack sources, such as independently operated, open-source-based attacks, will remain difficult to observe no matter the scheme, a well-constructed membership could provide meaningfully robust visibility. ACE membership should strive to include:

- **Model Providers**: Frontier models may serve as the reasoning layer for complex attacks or as capability augmentation for smaller or open-source-based attacks. Smaller and behind-the-frontier model providers must also be included. As they improve, smaller systems may also originate attacks, serve as components in multi-model attack chains, or become targets of agent hijacking.
- **Cloud Infrastructure Providers:** These providers supplement model-provider visibility, helping detect large-scale malicious inference activity, open-source malicious cloud deployments, and anomalous compute patterns associated with offensive cyber agent operations. If the ACE takes on a disruption coordination role, cloud providers would also be uniquely positioned to terminate malicious compute workloads.
- **Transnational Participation**: Global representation is required, as agentic attack infrastructure will span jurisdictions, cloud regions, and software ecosystems across borders. The GSE and several existing ISACs demonstrate that global institutional collaboration is feasible. The Financial Services ISAC (FS-ISAC), for example, includes institutions from over seventy countries.[165]

[164] CERT-EU, "The Cybersecurity Service for the Union institutions, bodies, offices and agencies."

Membership breadth must balance against a countervailing requirement: trust and exclusivity. Cyber threat intelligence research consistently identifies trust among participants as one of the most important enablers of effective information sharing.[166] A curated high-trust membership can improve data quality and provenance, simplify the negotiation of data-sharing agreements, reduce free-riding, and strengthen the institutional cohesion required for coordinated response. One possible solution is a tiered structure, including a smaller subgroup of systemically important members. A useful model is the FS-ISAC's Financial Systemic Analysis and Resilience Center (FSARC),[167] a select membership tier of the largest financial providers designed for deep systemic coordination.[168]

While national CERTs, sectoral ISACs, and other cybersecurity institutions reside outside the defined ACE membership, the ACE would seek to build a symbiotic coordinating relationship with these institutions. These existing institutions would ideally provide the ACE sector-specific indicators that can be mixed with ACE member data to aid agent detection and threat mitigation. Serving its public security mandate, the ACE would likewise lean on these institutions as distribution channels to ensure rapid diffusion of alerts and to coordinate disruption as relevant.

## Information Sharing and Analysis Infrastructure

Agentic threats will operate at machine speed that ACE information sharing infrastructure must be capable of matching. Further, if operations are decomposed across models, platforms, and vendors, detection cannot happen unless raw signals are shared in near real time to enable actors to draw the connections needed to identify threats. This may require real-time sharing of raw telemetry.

The E-ISAC's CRISP program offers a promising structural model for such real-time telemetry sharing. Under CRISP, participating electric utilities deploy a "passive information sharing device" that automatically transmits raw network activity telemetry to a centralized, sector-wide threat analysis body in near real time.[169] This allows the energy sector to rapidly spot sectoral threats.

---

[165] FS-ISAC, "FAQs."

[166] Wagner et al., "Cyber Threat Intelligence Sharing: Survey and Research Directions."

[167] Atkins and Lawson, "Cooperation amidst competition: cybersecurity partnership in the US financial services sector."

[168] The FSARC model also has value for disruption. Given its curated size, the FSARC has been able to effectively organize to actively collaborate with U.S. cyber command to disrupt cyberattacks on the financial sector (Bing, "Inside 'Project Indigo,' the quiet info-sharing program between banks and U.S. Cyber Command.").

[169] The program now covers approximately 70% of the North American electricity market, demonstrating that automated data sharing can operate at scale when supported by appropriate technical and governance structures. U.S. Department of Energy, "Energy Sector Cybersecurity Preparedness."

An ACE-equivalent would serve a slightly different purpose. Whereas CRISP detects attacks *against* member utilities, an ACE analog would need to identify malicious agents operating *through or across* member platforms—as attack vectors, command-and-control relays, or sources of coordinated malicious behavior. To operationalize this, the ACE can follow the E-ISAC's model and install passive behavioral sensors within member API gateways to monitor for agentic signatures, such as orchestration patterns or anomalous tool-chaining, or other signatures. Alternatively, members could use a federated network of agent honeypots (see Section 5: Agent Honeypots) distributed across participating platforms, with telemetry centrally analyzed. Either model naturally raises privacy, legal, and competitive concerns that would require governance and anonymization frameworks analogous to those underpinning CRISP itself. CRISP's significant success and global membership, however, suggests these challenges are surmountable.

The ACE's analytical demands may be substantial and must also be considered when provisioning resources and infrastructure. Existing cyber anomaly-detection systems already require significant computational resources,[170] and the adaptive, machine-speed character of agentic threats will further increase analytical complexity. Detection capabilities may need to analyze patterns such as agent orchestration behavior, tool-chain usage across platforms, model-switching patterns, and long-horizon task decomposition associated with autonomous agents. Meeting these demands will likely require advanced AI-assisted analytics, significant compute resources, and substantial data storage capacity. As has been the case with the Global Signal Exchange, which uses Google's Cloud Platform to facilitate sharing and AI analysis,[171] ACE members would be well-suited to meeting these technical requirements, potentially reducing operating overhead.

## Signal Correlation Mechanisms

Any ecosystem layer detection managed by the ACE will be improved by the ability to correlate signals across platforms and infrastructure layers. Two correlation challenges merit particular attention. The first challenge is correlating activity detected across the ecosystem with the agents responsible. If an agent interacts with two separate platforms, those platforms have few means to correlate those events or identify the agent responsible. The agent identity infrastructure discussed in previous sections (see: "Section 3: Agent Identity for Critical Infrastructure") would provide a baseline mechanism for associating activity detected across platforms with known agents. For detected, yet unidentified agents, behavioral fingerprinting techniques may be required to correlate activity within identity gaps.

The second challenge is correlating agent outputs. Malicious agent campaigns may fragment activity into benign-appearing subtasks distributed across multiple systems. Detection capabilities may therefore also benefit from analytical tools capable of tracing outputs across platforms and linking them to a shared origin. Both challenges represent active research frontiers. The ACE's institutional design should therefore anticipate ongoing investment in correlation capabilities.

---

[170] Campazas-Vega et al., "Malicious traffic detection on sampled network flow data with novelty-detection-based models."
[171] Storey and Zebarjadi, "The new Global Signal Exchange will help fight scams and fraud."

## Privacy, Security, and Legal Safeguards

A security compromise of ACE systems could expose sensitive data, undermine member trust, and potentially reveal the ecosystem's detection capabilities to adversaries. Robust physical and digital security measures are, therefore, foundational. Equally important is the protection of user privacy. Privacy concerns will also be paramount as detection-relevant data may include chat logs, reasoning traces, model interaction records, or other datasets that may be considered regulated, personally identifiable information.[172] Mishandling such information could expose member organizations to legal and reputational risk, therefore, privacy-preserving computation techniques, including federated analysis,[173] secure multi-party computation,[174] and homomorphic encryption,[175] should be explored to enable analysis while minimizing direct data sharing. Finally, legal risks must be mitigated. Organizations may be reluctant to share telemetry without clear liability protections or safe-harbor provisions governing cybersecurity information sharing. Policymakers should evaluate whether existing legal frameworks are sufficient to support ACE participation or whether additional protections will be required.

## Limitations and Uncertainties

Three major uncertainties are worth highlighting. Permissive legal frameworks are an essential prerequisite for success. Without the certainty that information can be shared without violating antitrust laws, privacy laws, trade agreements, and other restrictions, organizations may be slow, reluctant, or in some instances unable, to collaborate.These legal headwinds are especially acute considering recent flux in information-sharing regulations, including the uncertainty over the future of CISA 2015 (the United States' keystone information-sharing statute), ongoing uncertainty in the legal interpretation of the European Union's GDPR, and the mounting barriers created by increasingly common data sovereignty rules. These legal risks haven't yet disabled information sharing; however, without greater legal certainty, ambitious action will be less likely.

A second uncertainty is the impact natural market disincentives will have on collaboration. These include the reputational risks of potential data breaches, operational overhead, the potential loss of proprietary information, and the general effort collaboration requires.

---

[172] OpenAI, "Introducing OpenAI for Healthcare."
[173] Mrabet, "TrustFed-CTI: A Trust-Aware Federated Learning Framework for Privacy-Preserving Cyber Threat Intelligence Sharing Across Distributed Organizations."
[174] Science Direct, "Secure Multiparty Computation."
[175] ISACA, "Exploring Practical Considerations and Applications for Privacy Enhancing Technologies."

A third uncertainty is what will be required if the ACE is to manage offensive cyber agent disruption. As disruption is beyond the scope of this detection report, further research should outline those requirements, focusing on both the legal and technical prerequisites.

## Recommendations

1. **Governments, in collaboration with industry, should develop ACE Implementation Plans.**

As discussed above, the United States is the most obvious potential public sponsor. To spark action, the United States Congress should commission an ACE implementation report modeled on the congressionally commissioned National Artificial Intelligence Research Resource implementation plan[176] and developed in close conjunction with industry. Middle powers should also consider leading action and commissioning similar implementation plans. Such investigations should provide answers to the design considerations listed above. Investigative priority should be placed on institutional design, funding and governance models, membership scope, and legal uncertainties.

2. **Governments, in collaboration with agent service providers, should assess the feasibility of a CRISP-like sensing architecture tailored to agentic threat detection.**

Two models should be considered: passive behavioral sensors embedded in member infrastructure and a federated agent honeypot network. This assessment should be structured as a pilot program involving a small cross-sector subset of members and should aim to surface technical, regulatory, and competitive obstacles before broader deployment. Attention should be placed on both developing required technology and whether CRISP's anonymization and cost-sharing agreements might be adapted.

3. **Philanthropists, research funders, industry, and governments should invest in ecosystem-wide information sharing and analysis research.**

To minimize noted privacy, legal, and security risks, priority should be placed on research into the minimum-viable signal-sharing needed for ecosystem-wide detection and analysis. This research will build on more general research into identifying agent signatures by specifically investigating the minimum variety, volume, and sources of data needed to robustly detect cyber offensive agents.

Further research should investigate signal correlation mechanisms, studying how an identified agent might be reliably fingerprinted, how disparate agent outputs might be linked to a common source, and the general utility of agent IDs as a signal correlation mechanism (see Section 3: Agent Identity for Critical Infrastructure). Recognizing that attackers might cycle agents and models, further research should seek to investigate means of fingerprinting attacker identity. Potential avenues include stylometric analysis of attacker prompts and research into identity signatures that may be leaked from model outputs.[177]

[176] National Artificial Intelligence Research Resource Task Force, "Strengthening and Democratizing the U.S. Artificial Intelligence Innovation Ecosystem."

[177]The text of input prompts could provide signatures that can help identify malicious users. Research suggests LLMs could enable "large-scale online deanonymization" based on stylometric analysis of user-written text data. Likewise, model outputs and usage behaviors may leak signatures on the controlling user's identity. It has been found that the specific combination of agents a user employs can be strongly correlated with features of that user's identity. Further research has found that Deepseek significantly alters outputs based on user identity. For any model that behaves in a similar fashion, these altered outputs could provide details on a user's identity. Naturally, such research into attacker identity signatures should heed the significant privacy risks any such technique could yield. While there may be law enforcement benefits, sometimes a stone is best left unturned. (sources: Lermen et al., "Large-scale online deanonymization with LLMs."; Zhang and Zhang, "Assessing Deanonymization Risks with Stylometry-Assisted LLM Agent."; Zhang et al., "Exposing LLM User Privacy via Traffic Fingerprint Analysis: A Study of Privacy Risks in LLM Agent Interactions."; Stein, "CrowdStrike Research: Security Flaws in DeepSeek-Generated Code Linked to Political Triggers.")

# Acknowledgments

We would like to thank the following individuals for their input and feedback on our research: Sam Boger, Asher Brass, Alan Chan, Raja Sekhar Rao Dheekonda, Soham Mehta, Amin Oueslati, and Oleg Shakirov.

We would also like to extend special thanks to the Hewlett Foundation for helping to fund this report; Miro Pluckebaum for supporting this collaboration; and Ding Yao Wan, Theresa Chew, and Sonya Chan from the Cyber Security Agency of Singapore (CSA) for their close review and support.

Any errors, omissions, or views expressed in this paper are solely those of the authors. Inclusion in the acknowledgements does not constitute endorsement of our findings or recommendations.

# Appendix

## I. Call-Out Box: Possible Approaches to Develop Signatures for Offensive Cyber Agents

### Developing Signatures for Offensive Cyber Agents

Determining what agent-specific behavioral signatures look like is an open research problem. With conventional malware families, defenders can study captured samples, reverse-engineer their code, and build detection rules from observed characteristics. With human threat actors, tradecraft can be characterized through years of accumulated incident response data. Advanced autonomous cyber agents present a different challenge: there is, as yet, limited empirical data on what their operations look like at the network and endpoint level. Progress on this problem could be made through the following approaches.

First, security researchers could deploy advanced autonomous cyber agents and their precursors in controlled offensive security scenarios to study what behavioral traces they leave. This would generate useful data on agent-produced artifacts and behaviors under realistic conditions and potentially provide data even before real-world harm occurs.[178]

Second, security researchers could work backward from the operational requirements of autonomous cyber agents (e.g., the need to chain multiple tools together, acquire and manage compute resources, query external APIs for reasoning support, or maintain persistent access) to hypothesize what observable patterns these requirements would produce on a network. For instance, the GTIG findings on PROMPTFLUX and LAMEHUG[179] already reveal a distinctive pattern: malware that makes outbound API calls to LLM endpoints during execution, a behavior that would be highly anomalous for most legitimate software and essentially nonexistent in traditional malware.

Third, analyzing real-world data from measures like agent honeypots or agent-oriented incident reports could help better characterize autonomous agent activity, leading to the development of effective detection signatures.

---

[178] Irregular, "The Next Generation of Cyber Evaluations."
[179] Hacıoğlu, "LameHug: The First Publicly Documented Case of a Malware Integrating a LLM."

## II. Examples of Traditional Honeypots

| Honeypot | Deployment Details | Use Case |
| --- | --- | --- |
| **Canary Tokens**[180] | Canary tokens are unique markers embedded inside ordinary-looking files, documents, URLs, database entries, or API keys. They are placed inside real production environments, wherever an attacker or malicious insider might look. | The purpose is breach detection, with canary tokens acting as a lightweight tripwire. When someone opens or accesses a canary token, it silently reports back to the defender, providing the intruder's IP address, the name of the token accessed, and the time of access. However, canary tokens provide only a basic alert and limited metadata; they do not capture detailed information about the attacker's broader methods or tools. They can be useful to detect lateral movement and insider threats. |
| **Thinkst Canary** | Thinkst Canaries are full honeypot devices placed inside an organization's internal network. Each one emulates a real system, like routers or Linux servers, and runs realistic services, e.g., file shares. | The purpose is to detect intruders who have already gained access to the internal network—either external attackers moving laterally after an initial breach, or malicious insiders. When someone interacts with a Canary, the device immediately alerts the security team. The strategic value is early warning of a breach already in progress. |
| **T-Pot (Telekom Security)**[181] | T-Pot is an open-source platform that bundles more than 20 different honeypot tools into a single deployment. This can imitate several networks services, e.g., SSH login prompts, email servers, network appliances, etc. T-Pot is typically installed on an internet-facing server or cloud instance. | T-Pot is designed to give defenders and researchers a broad view of the threat landscape. Because it emulates many services at once, it captures a wide cross-section of attack types, from brute-force login attempts to malware delivery to web application exploits. The main advantage is breadth: a single deployment covers many attack vectors. It is used by security operations teams for monitoring and by researchers for studying attack trends. |
| **High-Interaction SSH Honeypot**[182] | A research honeypot running a genuine SSH service, deployed on the open internet for over a year. Unlike low-interaction honeypots that only emulate a login prompt, this allowed attackers who guessed valid credentials to log in and operate | The purpose was academic research into how human attackers actually behave after compromising a machine. The researchers studied the two-phase attack process they observed: first, automated brute-force password guessing, and then, for those who succeeded, hands-on intrusion, e.g., exploring the system, |

[180] Acalvio Technologies, "Canary Token."
[181] Telekom Security, "Introduction into T-Pot: A Multi-Honeypot Platform."
[182] Nicomette et al., "Set-up and deployment of a high-interaction honeypot: experiment and lessons learned."

| | | |
|---|---|---|
| | freely, but within a monitored environment. | downloading attack tools, and attempts to expand access. |
| **AWS Madpot**[183] | MadPot is a globally distributed network of honeypot sensors operated by AWS, with automated response capabilities. It leverages AWS's position as a major cloud provider to deploy tens of thousands of sensors that emulate hundreds of different services. When attackers deliver malware to a MadPot sensor, AWS executes the malicious code in an isolated sandbox environment to extract indicators of compromise from its behavior. | MadPot serves two purposes: gathering threat intelligence and actively disrupting attacks. Within 30 minutes of identifying a new threat, the system can automatically translate its findings into protective rules for AWS security services like GuardDuty and Network Firewall. AWS also shares findings with external parties, including government agencies, internet service providers, and domain registrars, to help disrupt malicious infrastructure beyond AWS's own network. |

## III. Additional Defensive Benefits of Agent Honeypots: Strengthening Defenses and Active Disruption

Similar to traditional honeypots, agent honeypots open up defensive options to slow down offensive agent operations, improve network defences against their attacks, and provide surface area to conduct counterattacks that could disrupt the agents operation.

**Identifying vulnerabilities and informing defensive priorities.** When an agent exploits a honeypot, defenders can learn which vulnerabilities it targeted and how. A sufficiently realistic honeypot environment that allows the agent to progress through multiple stages of an attack can reveal the full attack chain, including the tools deployed, the order in which access was escalated, and how the agent responded when it encountered obstacles. This allows organizations to harden the exploited vulnerabilities in their real systems and informs them about which threats they should prioritize.

**Slowing, distracting, and deterring attackers.** Time and compute that an agent spends interacting with a decoy resource is not spent on real systems. By designing honeypots to intentionally waste the agents time, honeypots can avert harm from valuable assets and slow down the attacker's progress.[184][185] Furthermore, the knowledge that honeypots might be present in a network should make attackers more hesitant to take offensive action.

[183] Amazon Staff, "Meet MadPot, a threat intelligence tool Amazon uses to protect customers from cybercrime."
[184] Ayzenshteyn et al., "Cloak, Honey, Trap: Proactive Defenses Against LLM Agents."
[185] Tatoris et al., "Trapping misbehaving bots in an AI Labyrinth."

**Enabling counterattacks and disruption.** Defenders can embed hidden instructions in honeypots that manipulate offensive AI agents. For example, prompt injection attacks can trick an agent into downloading and running code from a defender-controlled server, potentially giving defenders access to the attacker's own system.[186][187]

## IV. Estimated Costs of Various Honeypots

| Honeypot | Description | Estimated One-Time Setup Cost (USD) | Estimated Annual Cost (USD) |
|---|---|---|---|
| **Canary Tokens**[188] | Decoy credentials and files for intrusion detection (honeytoken) | ~$50 | ~$1.2K |
| **Thinkst Canary** | Decoy servers, routers, and devices to detect post-compromise activity (low-interaction honeypot, pre-built) | ~$600 | ~$4K–$5K |
| **T-Pot (Telekom Security)**[189] | Internet-facing bundle of decoy network services (low-interaction, pre-built) | ~$300–$2.2K | ~$13K–$14K |
| **High-Interaction SSH Honeypot**[190] | Full Linux OS to study post-compromise behaviors (high-interaction, self-built) | ~$67K–$115K | ~$61K–$120K |
| **AWS Madpot**[191] | 10,000s of sensors mimicking AWS workloads (enterprise-wide) | $5M-$15M/year | ~$11M–$30M |
| **Palisade LLM Honeypot** | SSH honeypot with LLM agent detection (low-interaction, agent honeypot) | ~$8K | ~$2K |

## V. Use Cases of Traditional Honeypots

Honeypots serve a range of defensive purposes. We categorize these into five uses:

---

[186] This counterattack was demonstrated in: Ayzenshteyn et al., "Cloak, Honey, Trap: Proactive Defenses Against LLM Agents." As well as in Heckel and Weller, "Countering autonomous cyber threats."

[187] It is unclear whether more sophisticated agents will remain vulnerable to such prompt injection attacks.

[188] Acalvio Technologies, "Canary Token."

[189] Telekom Security, "Introduction into T-Pot: A Multi-Honeypot Platform."

[190] Nicomette et al., "Set-up and deployment of a high-interaction honeypot: experiment and lessons learned."

[191] Amazon Staff, "Meet MadPot, a threat intelligence tool Amazon uses to protect customers from cybercrime."

1. **Threat Intelligence.** Honeypots can reveal what attackers are doing, how they operate, and who they are. This provides defenders an overview of the threat landscape and emerging trends. For example, T-Pot reveals trends in scanning behavior, vulnerability exploitation, and attack patterns.[192] Other honeypots log interactions in full detail and capture artifacts used in attacks. This reveals attackers' tactics, techniques, and procedures (TTPs) on specific targets, including novel strategies and exploits. Additionally, the gathered information can profile and attribute threat actors.[193] AWS's MadPot system, consisting of tens of thousands of honeypots, has uncovered and attributed multiple previously unknown advanced persistent threat (APT) campaigns such as Volt Typhoon.[194]

2. **Detection and Incident Response.** Because honeypots have no production value, any interaction with them is inherently suspicious. Thus, they provide intrusion alerts with almost no false positives. Canarytokens are the simplest example: fake credentials, API tokens, or documents sprinkled throughout an organization's infrastructure that trigger an immediate alert when accessed.[195] Grafana Labs used canarytokens to detect a real supply-chain attack. When an attacker exploited a vulnerable GitHub Action and tried to verify stolen AWS tokens, the canary token fired immediately.[196] Inside networks, devices like Thinkst Canary alert defenders to lateral movement and post-compromise reconnaissance.[197]

3. **Improving Defenses.** Intelligence gathered by honeypots can feed directly into defensive systems. MadPot's findings automatically update AWS GuardDuty detection rules and Network Firewall protections within 30 minutes of discovery. More generally, honeypots yield indicators of compromise (malicious IPs, malware hashes, credential dictionaries) that can be used to update firewall rules, IDPS signatures, and SIEM alert configurations. Nicomette et al.[198] used their honeypot to test whether specific defenses were sufficient or whether attackers would find workarounds. Honeypot data can also reveal which credentials have leaked and need to be rotated, and which attack vectors are being actively exploited against an organization's specific infrastructure.

4. **Active Disruption.** Honeypots can enable defenders to take offensive action against attacker infrastructure. AWS used MadPot intelligence to stop 1.3 million outbound botnet-driven DDoS attacks in Q1 2023 alone, and shared nearly a thousand C2 hosts with hosting providers—in one case, mapping a botnet's C2 infrastructure and working with the host to take it down within 48 hours. More broadly, intelligence from honeypots can be shared with law enforcement, CERTs, and other organizations to coordinate takedowns of phishing sites, C2 servers, and compromised infrastructure. This use requires honeypots that capture enough detail about attacker infrastructure (IPs, C2 protocols, malware samples) to enable action.

---

[192] GitHub, "tpotce."
[193] Nicomette et al., "Set-up and deployment of a high-interaction honeypot: experiment and lessons learned."
[194] Amazon Staff, "Meet MadPot, a threat intelligence tool Amazon uses to protect customers from cybercrime."; Ryland, "How AWS threat intelligence deters threat actors."
[195] Canary Tokens, "Create a Canarytoken. Deploy it somewhere."
[196] Moradian, "Canary tokens: Learn all about the unsung heroes of security at Grafana Labs."
[197] Thinkst Canary, "Canary Tools."
[198] Nicomette et al., "Set-up and deployment of a high-interaction honeypot: experiment and lessons learned."

5. **Slowing, distracting, and deterring attackers.** Honeypots waste attackers' time and divert them from real assets. In the Tularosa study,[199] 130 professional red teamers attacked a network containing decoy systems alongside real ones. 52% of attacker commands targeted decoys, reducing traffic to real assets by 25%. Only 1 out of ~60 participants correctly identified any decoy as fake. Crucially, participants who were merely *told* that deception might be present—even without encountering it—were slower, made less progress, and triggered more decoys. This deterrence effect operates even when no honeypots are actually deployed: the mere possibility of deception forces attackers to spend resources verifying targets. Ferguson-Walter et al. found that when attackers suspect deception, 38% increase activity (potentially exposing themselves further), while 21% decrease it, 18% avoid suspicious targets, and 10% pause entirely.[200]

**Enabling counterattacks and disruption:** Actively interfering with hostile agents through prompt injection and other manipulation techniques, potentially redirecting agent behavior, planting false information, or halting unauthorized agent activity.

**Identifying vulnerabilities and informing defensive priorities.** When an agent exploits a honeypot, defenders can learn which vulnerabilities it targeted and how. A sufficiently realistic honeypot environment that allows the agent to progress through multiple stages of an attack can reveal the full attack chain, including the tools deployed, the order in which access was escalated, and how the agent responded when it encountered obstacles. This lets security teams prioritize threats and patch or harden the same vulnerabilities in production systems before the same exploits are used there.

# VI. Design Space of Agent Honeypots

Honeypots vary widely in their design and deployment. Nawrocki et al. identify several axes along which they differ:[201]

- **Interaction depth:** Low-interaction honeypots emulate only a small set of services and are mainly used to gather statistics, while high-interaction honeypots provide real operating systems and capture detailed attacker behavior at the cost of greater complexity and risk of compromise.

---

[199] Ferguson-Walter et al., "Examining the Efficacy of Decoy-based and Psychological Cyber Deception."
[200] Ferguson-Walter et al., "Cyber expert feedback: Experiences, expectations, and opinions about cyber deception."
[201] Nawrocki et al., "A Survey on Honeypot Software and Data Analysis."

- **Production vs. research:** Production honeypots are deployed within an organization's network to improve its security, while research honeypots are operated by researchers to study threats and develop general countermeasures.
- **Server vs. client:** Server honeypots wait passively for attackers to connect, while client honeypots actively visit potentially malicious resources such as websites.
- **Physical vs. virtual:** Honeypots can run on dedicated hardware or as virtualized instances on a host machine. Virtualization enables many honeypots to run on a single host but introduces detectable artifacts.

Beyond these, honeypots also differ in:

- **Simulated services:** From SSH to cloud workloads, which determine what kinds of attacks they attract.
- **Scope:** From a single honeypot emulating one service to honeynets that simulate entire networks.
- **Placement:** Internet-facing honeypots receive large volumes of opportunistic attacks, while internally deployed honeypots see little traffic but generate high-confidence alerts about post-compromise lateral movement.
- **Heterogeneity:** Deploying different types of honeypots covers a wider range of attacks.[202]
- **Dynamism:** While static honeypots always act the same, dynamic honeypots adapt to changes in network status and attacker behavior.[203]

# VII. More Detailed Explanation of Different Potential Honeypot Developers and Deployers

Agent honeypots could be developed and deployed by a range of actors, each with different capabilities, incentives, and threat models. Some are well-positioned to build novel honeypot designs; others are better suited to deploy and maintain them at scale. In many cases, the most effective approach will involve partnerships between developers and deployers.

**Frontier AI companies** are uniquely positioned to deploy honeypots against threats from their own internally deployed models. A highly capable, misaligned model might attempt to exfiltrate its own weights, escalate its privileges, or undermine oversight mechanisms. Honeypots placed inside the company's internal systems, such as fake model weights, honeytoken credentials, or decoy infrastructure, could catch such attempts, alerting security teams and providing direct evidence about the model's behavior and alignment. Beyond this internal threat model, frontier labs also control assets (model weights, training infrastructure, research data) that make them attractive targets for external agent-driven attacks, and they have the technical sophistication to build and maintain realistic deception environments.

[202] Javadpour et al., "A comprehensive survey on cyber deception techniques to improve honeypot performance."
[203] Javadpour et al., "A comprehensive survey on cyber deception techniques to improve honeypot performance."

**Cloud compute providers** are natural deployers of agent honeypots. Their infrastructure is a high-value target for a range of threats: attackers using agents to compromise customer workloads, steal credentials, or establish persistent access, as well as (more speculatively) rogue agents seeking compute to sustain their own operations. Providers like AWS (which already operates the MadPot honeypot network) have the infrastructure to deploy honeypots at massive scale, the technical expertise to build convincing emulations of their own services, and the commercial incentive to understand the threats facing their networks and customers.

**Critical infrastructure operators** face a distinctive threat: AI agents targeting industrial control systems, energy grids, water treatment facilities, or transportation networks. These operators often have limited in-house capability to build sophisticated deception environments, but they control the network segments where industrial control system (ICS) honeypots need to be deployed.

**Financial institutions and major banks.** They are among the most heavily targeted organizations for cyberattacks, already invest heavily in threat intelligence and deception technology, and have the security budgets and in-house expertise to deploy sophisticated honeypots. They also face distinctive threats (fraud, theft, market manipulation) that could attract agents with specific objectives.

**Security researchers and threat intelligence companies** have the expertise to develop novel honeypot designs and the professional motivation to be first to identify emerging threats. Threat intelligence firms can deploy research honeypots broadly across the internet to gather data on the evolving threat landscape, and package findings into products for their clients, creating a direct commercial incentive to invest early. Academic researchers can contribute by developing and open-sourcing new agent-specific detection techniques, as the Palisade Research honeypots project has demonstrated.

**AI safety researchers** are interested in questions about AI-driven cyberoffense that go beyond traditional cybersecurity: how are threat models involving autonomous agents actually manifesting in the real world? What are the offensive capabilities of models deployed in uncontrolled settings? Agent honeypots could provide empirical evidence on both questions, informing risk assessments at AI companies, guiding deployment decisions, and shaping policy responses. They could also serve as one of the few early warning systems for the emergence of rogue agents operating outside human control.

**Intelligence and national security agencies** are already major producers and consumers of cyber threat intelligence, and AI-driven cyberattacks are a direct national security concern, particularly when adversary nations deploy agents as offensive tools. Agencies have the resources, technical capability, and mandate to deploy large-scale honeypot networks, and the intelligence gathered would feed into broader assessments of how foreign actors are incorporating AI into their cyber operations. They could also fund and coordinate honeypot programs that individual organizations lack the resources to operate alone, particularly in critical infrastructure sectors.

**Existing honeypot providers** already specialize in building and selling deception technology to network defenders. Adapting existing honeypot products to incorporate agent detection capabilities represents a natural extension of their current offerings, and could be the fastest path to broad deployment since it builds on established distribution channels and customer relationships.

## VIII. A Note on the Proposed United States AI-ISAC

In 2025, the United States AI Action Plan proposed am AI-ISAC led by the Department of Homeland Security (DHS) "to promote the sharing of AI-security threat information and intelligence across U.S. critical infrastructure sectors."[204] As this stated goal appears to more closely mirror the public security orientation of an ACE over the traditional sectoral security focus of ISACs, the effort, if it gains momentum, could provide an institutional basis for the creation of an ACE. This U.S.-specific path may be especially worthwhile if cross-border legal complexities challenge coordinating a global ACE.

That said, an AI-ISAC led by the United States or any single nation will face challenges when trying to maximize the potential of AI threat information sharing, analysis, and disruption coordination. First, if an AI-ISAC matches the broad security mandate of existing ISACs, which often extends beyond cyber risks, its resources and focus may be spread too thin to match the task at hand. Second, like many government initiatives, a government-run AI-ISAC would need to navigate fiscal uncertainties that could affect the sustained resources this work will require. Third, the privacy concerns that come with government-led programs could deter the sharing of essential, yet privacy-sensitive data such as chatlogs or reasoning traces. These challenges could deeply hamstring effectiveness. Finally, as agent security risks and the data required to detect them will not be confined by borders, any domestically scoped threat-sharing program may lack the threat environment visibility that success requires.

[204] The White House, "Winning the Race: America's AI Action Plan."